\documentclass[12pt,preprint]{aastex}
\usepackage{graphicx}
\usepackage{subfigure}
\usepackage{color}
\usepackage{natbib}  

\begin{document}

\title{\textit{XMM-Newton} Observations of a Complete Sample of Optically Selected Type 2 Seyfert Galaxies}

\author{Stephanie M. LaMassa$^{1}$, Timothy M. Heckman$^{1}$, Andrew Ptak$^{1}$, Ann Hornschemeier$^2$, Lucimara Martins$^3$, Paule Sonnentrucker$^1$, and Christy Tremonti$^4$}

\affil{$^1$The Johns
  Hopkins University, $^2$NASA Goddard Space Flight Center,$^3$ NAT - Universidade Cruzeiro do Sul, $^4$Max-Planck Institute for Astronomy}

\begin{abstract}

The majority of Active Galactic Nuclei (AGN) suffer from significant obscuration by surrounding dust and gas. The penetrating power and sensitivity of hard X-ray observations allows obscured AGN to be probed out to high redshifts. However, X-ray surveys in the 2-10 keV band will miss the most heavily-obscured AGN in which the absorbing column density exceeds $\sim10^{24}$cm$^{-2}$ (the Compton-thick AGN). It is therefore vital to know the fraction of AGN that are missed in such X-rays surveys and to determine if these AGN represent some distinct population in terms of the fundamental properties of AGN and/or their host galaxies. In this paper we present the analysis of \textit{XMM-Newton} X-ray data for a complete sample of 17 low-redshift Type 2 Seyfert galaxies chosen from the Sloan Digital Sky Survey based solely on the high observed flux of the [OIII]$\lambda$5007 emission-line. This line is formed in the Narrow Line Region hundreds of parsecs away from the central engine. Thus, unlike the X-ray emission, it is not affected by obscuration due to the torus surrounding the black hole. It therefore provides a useful isotropic indicator of the AGN luminosity. As additional indicators of the intrinsic AGN luminosity, we use the Spitzer Space Telescope to measure the luminosities of the mid-infrared continuum and the [OIV]25.89$\mu$m narrow emission-line. We then use the ratio of the 2-10 keV X-ray luminosity to the [OIII], [OIV], and mid-infrared luminosities to assess the amount of X-ray obscuration and to distinguish between Compton-thick and Compton-thin objects. The various diagnostics of AGN luminosity with heavily obscured hard X-ray emission are in broad agreement. We find that the majority of the sources suffer significant amounts of obscuration: the observed 2-10 keV emission is depressed by more than an order-of-magnitude in 11 of the 17 cases (as expected for Compton-thick sources). Thus, surveys in the rest-frame 2-10 keV band will be significantly incomplete for obscured AGN. We find a strong inverse correlation between the ratio of the 2-10 keV X-ray and [OIII] (or [OIV] or mid-IR) fluxes and the equivalent width of the 6.4 keV Fe K$\alpha$ line. This demonstrates that the weak hard X-ray continuum emission is due to obscuration (rather than to intrinsically weak emission). In several cases the large amount of obscuration is not consistent with the values of absorbing column density derived from simple spectral fits to the data. We run simulations of a more physically realistic model with partial covering of the X-ray source plus Compton scattering, and show that such models are consistent with the data. We show that the distribution of obscuration in the 2-10 keV band in our sample is similar to what is seen in samples selected in the {\it Swift} BAT energy-band (14-195 keV). This implies that the BAT surveys do recover a significant fraction of the local population of Compton-thick AGN. Finally, we find no relationship between the amount of X-ray obscuration and the other properties of the AGN and its host galaxy. Hence, Compton-thick and Compton-thin sources do not seem to trace distinct populations.

\end{abstract}


\section{Introduction}

In the standard model (e.g. Urry \& Padovani 1995), Active Galactic Nuclei (AGN) consist of an accretion disk around a supermassive black hole (SMBH) surrounded by a torus of dust and gas. The accretion disk is a source of ultraviolet and optical continuum emission, while X-rays are believed to originate through inverse Compton scattering of soft seed photons from the disk by hot electrons in a surrounding corona. Dust in the torus absorbs the emission from the accretion disk and re-emits it in the mid-infrared. Type 1 (unobscured) and Type 2 (obscured) AGN are intrinsically the same, but they display different characteristics because of their orientation. In Type 2 AGN the torus is viewed nearer edge-on, obscuring the central source, whereas in Type 1 AGN, the torus is viewed nearer face-on, providing an unobscured view of the central source \citep{Anton}. Obscured AGN have long been identified by their narrow high-ionization emission lines in the optical band, produced by photoionized gas far above and below the plane of the torus. Obscured AGN can also be recognized by the mid-infrared emission from the torus. Optical and infrared surveys show that obscured AGN constitute the majority of the population (e.g. Risaliti et al. 1999; Hao et al. 2005).

Hard X-rays (E $>$ 2 keV) also offer a potentially powerful way to find obscured AGN. So long as the obscuring column density (N$_{H}$) is $< 10^{24}$ cm$^{-2}$ (the Compton-thin regime), hard X-rays will suffer little absorption. However, observations of a local ($<$0.035) sample of 49 Type 2 AGN show that in fact many of these are very weak in the 2-10 keV X-ray band (Guainazzi et al. 2005). In the Chandra Deep Field South, only 82 out of over 300 AGN with optical spectroscopy were ``X-ray bright'' ($>$80 counts in the 2-10 keV band). Though this sample includes AGN up to much higher redshifts (up to $z \sim 3$), the subsamples of X-ray bright and X-ray faint targets have similar redshift distributions, indicating that the X-ray faint nature of the sample is not due to selection effects but rather that the majority of targets are weak X-ray emitters (Tozzi et al. 2006). The cases where the X-ray emission is severely attenuated are interpreted as cases in which the torus is Compton-thick (N$_{H} >$ 10$^{24}$ cm$^{-2}$). More recent surveys with {\it INTEGRAL} and {\it Swift} suggest that a at least some low redshift ($z <$ 0.05) Type 2 AGN remain weak emitters even at substantially higher photon energies ($>$ 10 keV, e.g. Treister et al. 2009 for $z <$ 0.05, but see Mel\'{e}ndez et al. 2008). This would require even higher obscuring column densities ($>10^{25}$ cm$^{2}$). 

How do we know that the weakness of the hard X-ray emission in these Type 2 AGN is due to obscuration rather than an intrinsic property (e.g. Brightman \& Nandra 2008)? Fortunately, there are signatures of obscuration in the X-ray spectra themselves. The weak X-ray emission that is seen in such cases often has a relatively flat X-ray spectrum and a strong Fe K$\alpha$ emission-line. These features are indicative of heavily reprocessed emission such as from reflection of hard X-rays off of the torus wall. This is as expected for a highly obscured X-ray source.

The weakness of the hard X-ray emission in many Type 2 AGN (regardless of its origin) implies that hard X-ray surveys do not fully probe the AGN population. This is important from several perspectives. First, phenomenological models often invoke a significant population of obscured AGN in order to successfully match the spectral energy distribution of the cosmic X-ray background at energies above 10 keV \citep{Comastri}. Are such models consistent with the observed properties of obscured AGN? Second, deep X-ray surveys can detect AGN out to high redshift. They then allow us to probe the cosmic evolution of the AGN population down to luminosities and SMBH masses well below the levels reached in optical surveys of quasars. Are the implications drawn from such surveys affected by a bias against Compton-thick AGN?

To address these issues, it is important to study a complete sample of AGN selected without regard to their hard X-ray properties. Optical spectroscopic surveys can in principle provide such a sample. Type 2 AGN in these surveys are identified by the high-excitation narrow emission from photoionized gas in the Narrow Line Region (NLR), material located hundreds of parsecs out along the polar axis of the torus. This emission is therefore unaffected by the obscuration of the torus itself and to first-order is radiated isotropically. Comparison of the luminosity of the gas in the NLR to that in hard X-rays can then be used to estimate the amount of X-ray obscuration. The most commonly used optical emission-line is [OIII]$\lambda$5007 (e.g. Bassani et al. 1999; Heckman et al. 2005). Several indicators of the intrinsic AGN luminosity also exist in the mid-infrared band where the effects of obscuration will be minimal. A close analog to the [OIII]$\lambda$5007 line is the luminosity of the [OIV] 25.89$\mu$m line from the NLR (e.g. Mel\'{e}ndez et al. 2008, Diamond-Stanic et al. 2009, Rigby et al. 2009). The luminosity of the mid-infrared continuum from the torus is also a useful indicator, although the torus is not believed to emit in a fully isotropic fashion (e.g. Pier \& Krolik 1993; Nenkova et al. 2008).

In this paper we present X-ray analysis of a nearly complete sample of 17 Type 2 Seyfert galaxies (Sy2) selected from the Sloan Digital Sky Survey (SDSS) based on their high [OIII]$\lambda$5007 flux. Using pointed \textit{XMM-Newton} observations of these sources, we derive their X-ray spectral characteristics and calculate their hard X-ray luminosities. We then use the SDSS data to measure the luminosities of [OIII]$\lambda$5007 lines and Spitzer Space Telescope Infrared Spectrograph (IRS) data to measure the luminosities of the mid-infrared continuum and of the [OIV] 25.89$\mu$ line. We use these data to assess the amount of X-ray obscuration. We also use measurements of the Fe K$\alpha$ line as an additional tool for probing highly obscured X-ray emission. Finally, we use the SDSS and Spitzer data to derive the basic intrinsic properties of the AGNs and their host galaxies. This allows us to determine whether the Compton-thick and Compton-thin populations differ in any systematic way. 

\section{XMM-Newton Observations and Data Reduction}

\subsection{Sample Selection}

Our Type 2 AGN are drawn from a parent sample of the spectra of roughly 480,000 galaxies in the SDSS Data Release 4 $=$ DR4 \citep{SDSS}. This parent sample of ``SDSS Main Galaxies'' \citep{Strauss} is complete over the SDSS DR4 footprint and is selected solely on the basis of the galaxy r-band apparent magnitude (14.5 $< r <$ 17.77). The diagnostic line ratio plot of [NII]/H$\alpha$ vs. [OIII]/H$\beta$ in Kauffmann et al. (2003 hereafter K03) and Kewley et al. (2006) was used to identify Type 2 AGN based on the SDSS spectra. Our sample was then defined as the 20 such objects having an observed [OIII]$\lambda$5007 flux above 4$\times 10^{-14}$ erg cm$^{-2}$ s$^{-1}$, corresponding to a luminosity of $\sim$10$^{^41}$ erg/s at the median distance for our sample ($\sim$150 Mpc). XMM-Newton data exist for 17 of the 20 objects in our sample, so the present sample is nearly complete.

We note that our sample is highly complementary to the sample of ``Type II Quasars'' investigated by Zakamska et al. (2003; 2008) and Ptak et al. (2006). These objects were selected on the basis of a high [OIII]$\lambda$5007 luminosity (rather than on the basis of a flux limit as we have done). The Type II Quasars were drawn from the full set of SDSS spectra, rather than from the homogeneous SDSS Main Galaxy sample. These objects define the extreme high luminosity end of the Type 2 AGN population, while our sample deals with typical Type 2 AGN.

\subsection{Data Analysis}

We were awarded \textit{XMM-Newton} time to observe 15 members of the 20 from this sample for a nominal 23 ks per target before filtering. Two additional targets from this sample were previously observed with \textit{XMM-Newton} and were added: 0325-0608 (Mrk 609, PI: Aschenbach) and 1218+4706 (PI: Page), bringing the sample total to 17. Table \ref{sample} lists the targets along with their ObsID number. The luminosity distances listed in Table \ref{sample} are based on the optical spectroscopic redshift, adopting a cosmology of H$_{o}$ = 71 km s$^{-1}$ Mpc$^{-1}$, $\Omega_{M}$ = 0.27 and $\Omega_{\Lambda}$ = 0.73.

The \textit{XMM-Newton} data were reduced using XAssist \citep{x_ref}, which runs the Science Analysis System (SAS) tasks to filter the raw data, generate light curves, clean the data for flaring, and extract spectra and associated response files (\textit{rmfs} and \textit{arfs}) for user-defined sources. Table \ref{sample} lists the good exposure time for each detector. For three sources, 1018+3722, 1111+0228, and 1437+3634, flaring was significant, so we applied minimal background filtering. The net X-ray counts, soft counts (0.5 - 2 keV) and hard counts (2 - 10 keV) are listed in Table~\ref{counts}.

For most of the sources, there were sufficient counts to group the PN, MOS1, and MOS2 data by $>$5 counts/bin (and in a majority of these cases, $>$10 counts/bin) and we used the $\chi^{2}$ statistic for the spectral analysis. The following targets were grouped by 2-3 counts/bin and were therefore analyzed using the C-statistic (XSpec handles spectral fitting via C-stat better when the data are slightly binned rather than unbinned, e.g. Teng et al 2005): 0053-0846, 1123+4703, and 1346+6423. One target, 1218+4706, fell on the chip gap in the PN detector, so we analyzed only the MOS1 and MOS2 data for this source.

The PN, MOS1 and MOS2 data were fit simultaneously for each source in XSpec, with all parameters tied together except for a constant multiplicative factor, which was frozen at 1 for the PN spectra and allowed to vary between 0.8 to 1.2 for MOS1 and MOS2, as the responses among detectors should generally not vary by greater than $\sim$20$\%$ (http://xmm.vilspa.esa.es/docs/documents/CAL-TN-0052.ps.gz); though in some cases, these limits were extended slightly to better fit the data when necessary (i.e. when the error on the statistical count rate is $>$20\%). We first fit all targets with a simple absorbed power law. The parameters derived from these fits are given in Table~\ref{pow}.

In many cases, a second power law component at higher energies was needed to accommodate the data, evident by both the shape of the spectrum and the inability of a single absorbed powerlaw to accurately model the data as reflected by the high $\chi^{2}$ value for a single power law fit. The targets best fit by a single absorbed power law are 0325-0608, 0959+1259, 1018+3722, 1123+4703, 1346+6423 and 1437+3634. The remaining 11 targets were fit by a double absorbed power law, where the photon indices for the 2 powerlaw components were tied together. In some cases (0325-0608, 0800+2636, 0824+2959, 1111+0228, 1123+4703 and 1218+4706), the best-fit absorption was the same as the Galactic absorption, so we froze this parameter to the Galactic value for these spectral fits; for 1346+6423, the spectral fit was best constrained by freezing absorption to the Galactic value and the photon index to 1.8. The best-fit parameters from these fits are listed in Table \ref{pow}, along with their corresponding $\chi^2$ values. As shown in Table \ref{ftest}, the double power law fit passes the F-test at almost or better than the 3$\sigma$ confidence level for all 11 targets to which it was applied, indicating that this model more accurately represents the data. Though applying the F-test to certain astrophysical tests is inappropriate (e.g. Protassov et al. 2002), the high probabilities derived indicate that the more complex models are likely statistically robust. This second power law component indicates that we are observing scattered and/or reflected AGN emission from these sources.

We note that several targets have photon indices quite a bit steeper than commonly observed in Seyfert galaxies, with a typical range being e.g. 0.5 $< \Gamma <$ 2.3 (Cappi et al. 2006): 0804+2345, 1018+3722, 1136+5657, 1147+5226, and 1157+5249\footnote{The lower error range for $\Gamma$ on these sources is greater than 2.5}. Better modeling of the soft emission, as discussed below, can result in a less steep photon index for the nuclear component.

Levenson et al. (2004; 2005) and Guainazzi et al. (2005) showed that soft thermal emission from hot gas associated with a starburst is commonly found in Seyfert galaxies. We will discuss the evidence for starbursts in our sample based on the SDSS and Spitzer IRS spectra in a future paper. Here we simply accommodate the possible presence of a starburst by fitting each of the targets with a thermal component (APEC in XSpec) added to the aforementioned absorbed single or double powerlaw model. The metal abundances were initially fixed to solar, allowing only the plasma temperature and normalization to be fit. We were able to fit the metal abundances for 2 of the sources: 0053-0846 and 1123-4703.  The fit parameters from the powerlaw plus thermal model are listed in Table \ref{apec}. For 4 of the targets (0325-0608, 1018+3722, 1123+4703, and 1147+5226), addition of the thermal component significantly improved the results at greater than the $3\sigma$ level according to the F-test (see Table \ref{ftest_apec}). We note that inclusion of this model component causes the power law index to decrease and to fall within the more commonly observed range for 1147+5226 and 1157+5249; though $\Gamma$ increases for 1018+3722, the lower end of the 90\% confidence level for this parameter now falls within the typical range observed in Seyferts. However, the power law indices remained too steep for 0804+2345 and 1136+5667 (4.94$^{+0.86}_{-0.95}$ and 3.11$^{+0.43}_{-0.35}$, respectively) and allowing the two photon indices to be fit independently caused more unphysical results: a negative $\Gamma_2$ for 0804+2345 and a much steeper $\Gamma_2$ for 1136+5667. Hence we froze $\Gamma_2$ at 1.8 for these sources.

Using the APEC model, we calculated the thermal X-ray luminosity (L$_{APEC}$) for all sources. We also fit these sources with the abundance of the thermal component frozen to 0.3 since though solar abundances would be expected in spiral galaxies, lower abundances are observed in practice when X-ray binaries are not resolved out, ``diluting'' the observed abundances (e.g. Ptak et al. 1999). We find the fit parameters to be consistent with those where the abundance is frozen to solar and find the thermal luminosity to be consistent between both cases, with at most a factor of two difference. In a future paper, we will examine the relationship between thermal X-ray emission and star formation rate (SFR).

\begin{figure}[f]
\centering
\subfigure[0800+2636]{\includegraphics[scale=0.30,angle=-90]{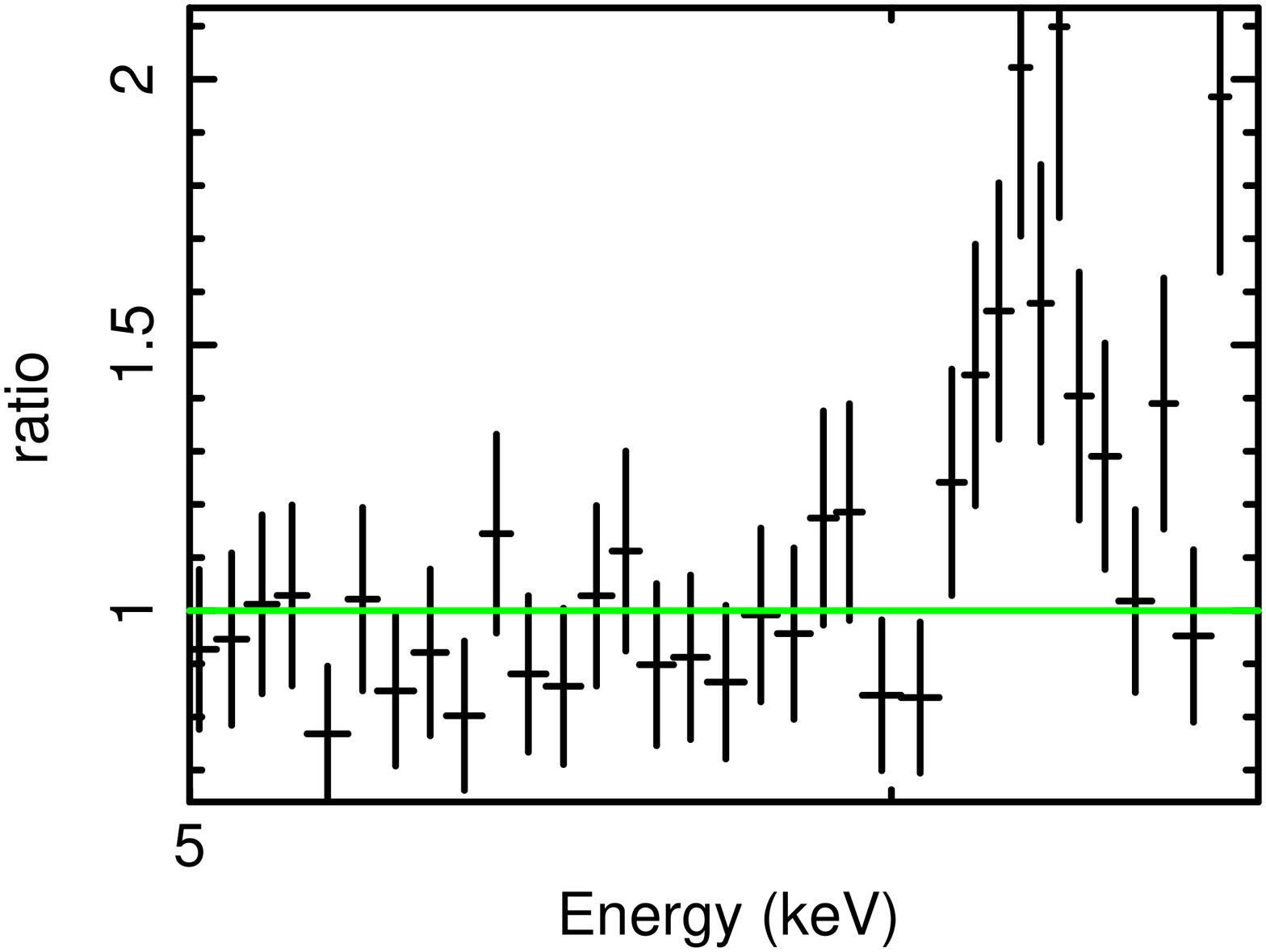}}
\subfigure[0824+2959]{\includegraphics[scale=0.30,angle=-90]{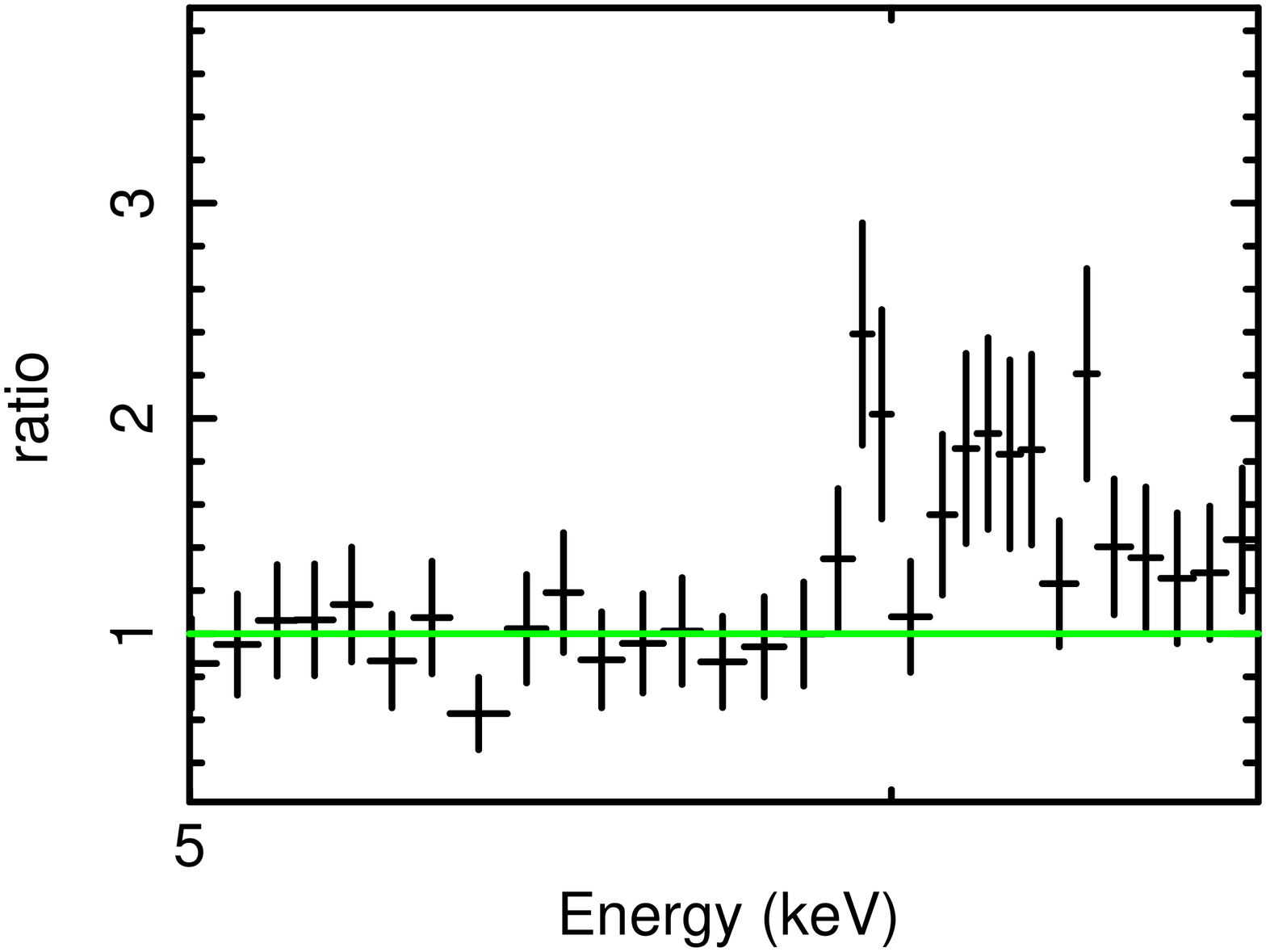}}
\subfigure[0959+1259]{\includegraphics[scale=0.30,angle=-90]{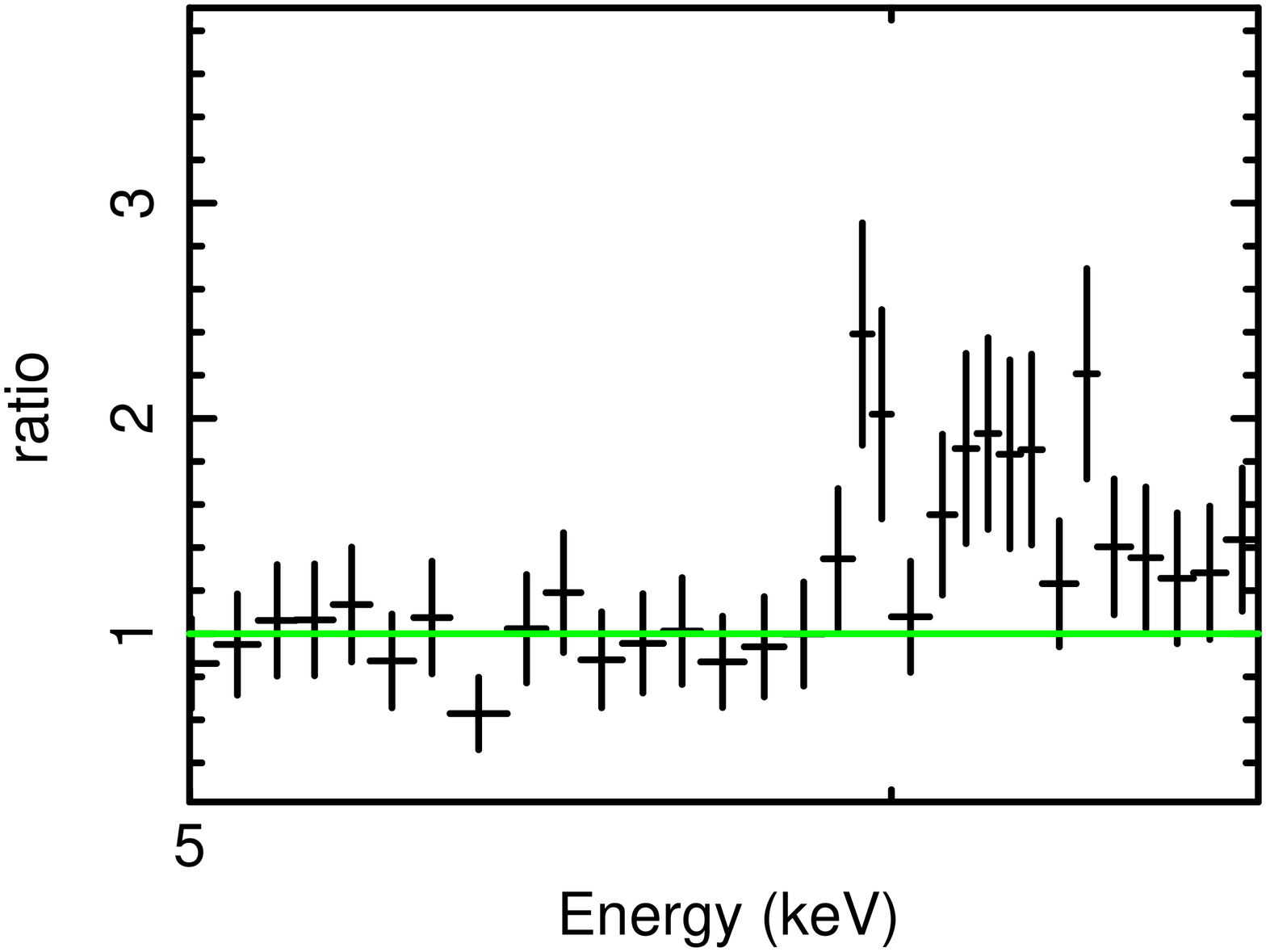}}
\subfigure[1238+0927]{\includegraphics[scale=0.30,angle=-90]{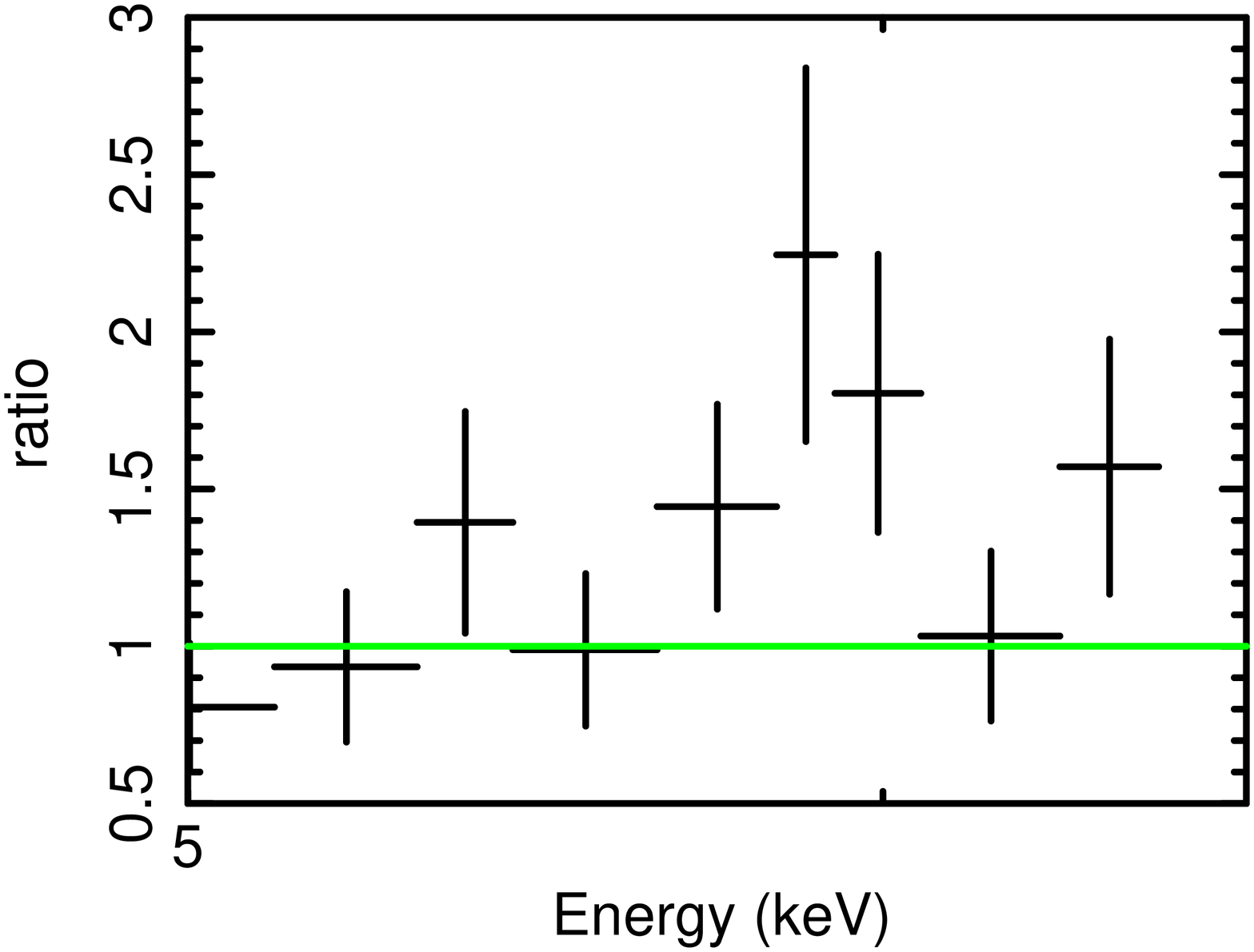}}
\subfigure[1323+4318]{\includegraphics[scale=0.30,angle=-90]{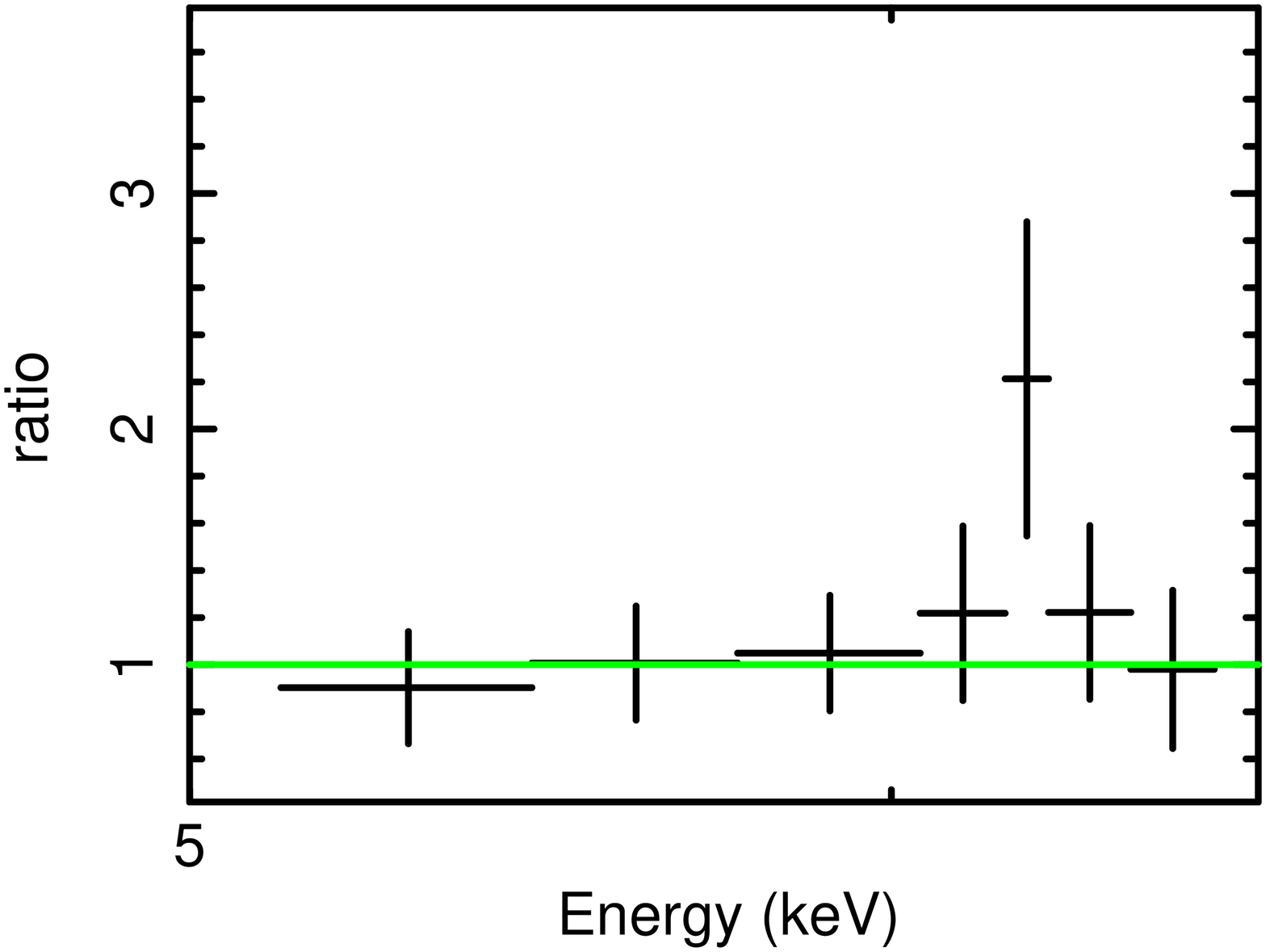}}
\caption{\label{ka_ratio} Data-to-model ratio plots of sources where the Fe K$\alpha$ line was detected at high significance for the 5 targets with the highest signal-to-noise ratio. In these plots, the spectra were fit with the best-fit model (double absorbed power law for 0800+2636, 0824+2959, 1238+0927 and 1323+4318 and single absorbed power law for 0959+1259) without the gaussian component added. The residuals at the Fe K$\alpha$ energy indicate that this line is present. }
\end{figure}

We detected the Fe K$\alpha$ emission-line in nine cases. For five cases with the highest signal-to-noise (0800+2636, 0824+2959, 0959+1259, 1238+0927 and 1323+4318) we were able to measure the energy and, for 4 of these, the width of the K$\alpha$ line while the other 4 detected sources were fit with a gaussian component frozen at the expected Fe K$\alpha$ line energy (i.e. 6.4/(1+z) keV) and a $\sigma$ of 0.01 keV\footnote{For the 5 cases with fitted Fe K$\alpha$ line energy, the fitted energy is consistent with the rest-frame Fe K$\alpha$ energy, 6.4 keV.}. To assess the significance of the Fe K$\alpha$ detection, we simulated 1000 spectra for each of the 9 sources based on the best-fit high energy (i.e. 4-8 keV) power law model. These simulated spectra were then fit with a power law and gaussian component to assess the probability that the Fe K$\alpha$ feature is detected due to random noise (we performed rigorous error analysis on a sub-set of these simulated spectra and found no statistical significant change in the Fe K$\alpha$ detection when new global minimizations were found). The simulation results indicate that the chance of detecting the Fe K$\alpha$ line at or above the observed value due to random variations is less than 1\% for 0800+2636, 0824+2959, 0959+1259, 1147+5226, 1157+5249, 1238+0927 and 1323+4318; less than 6\% for 0325-0608, and less than 14\% for 1218+4706. Data-to-model ratio plots for the 5 cases with the highest signal to noise are shown in Figure~\ref{ka_ratio}, where the spectra were fit without the line component; the presence of residuals at the Fe K$\alpha$ energy indicate that this line is present. For the 8 undetected targets, we were only able to obtain 3-$\sigma$ upper limits on the flux. The measurements for all 17 sources are listed in Table \ref{Fe}, where the Fe K$\alpha$ luminosities and EWs are the observed values. 

The 5 targets for which we were able to measure the energy of the Fe K$\alpha$ line were also fit with a \textit{diskline} model \citep{Fabian} as this model applies to Fe K$\alpha$ emission resulting from reflection of photons off the accretion disk. The results are listed in Table \ref{disk}. The inclination angle was constrained for 3 targets (0800+2636, 0824+2959, and 1238+0927), indicating that these lines are likely Doppler broadened. The EWs derived from the \textit{diskline} model are listed in Table \ref{Fe}. However, this model does not provide an improved fit over the use of a gaussian component to model the Fe K$\alpha$ line.

\begin{figure}[f]
\plotone{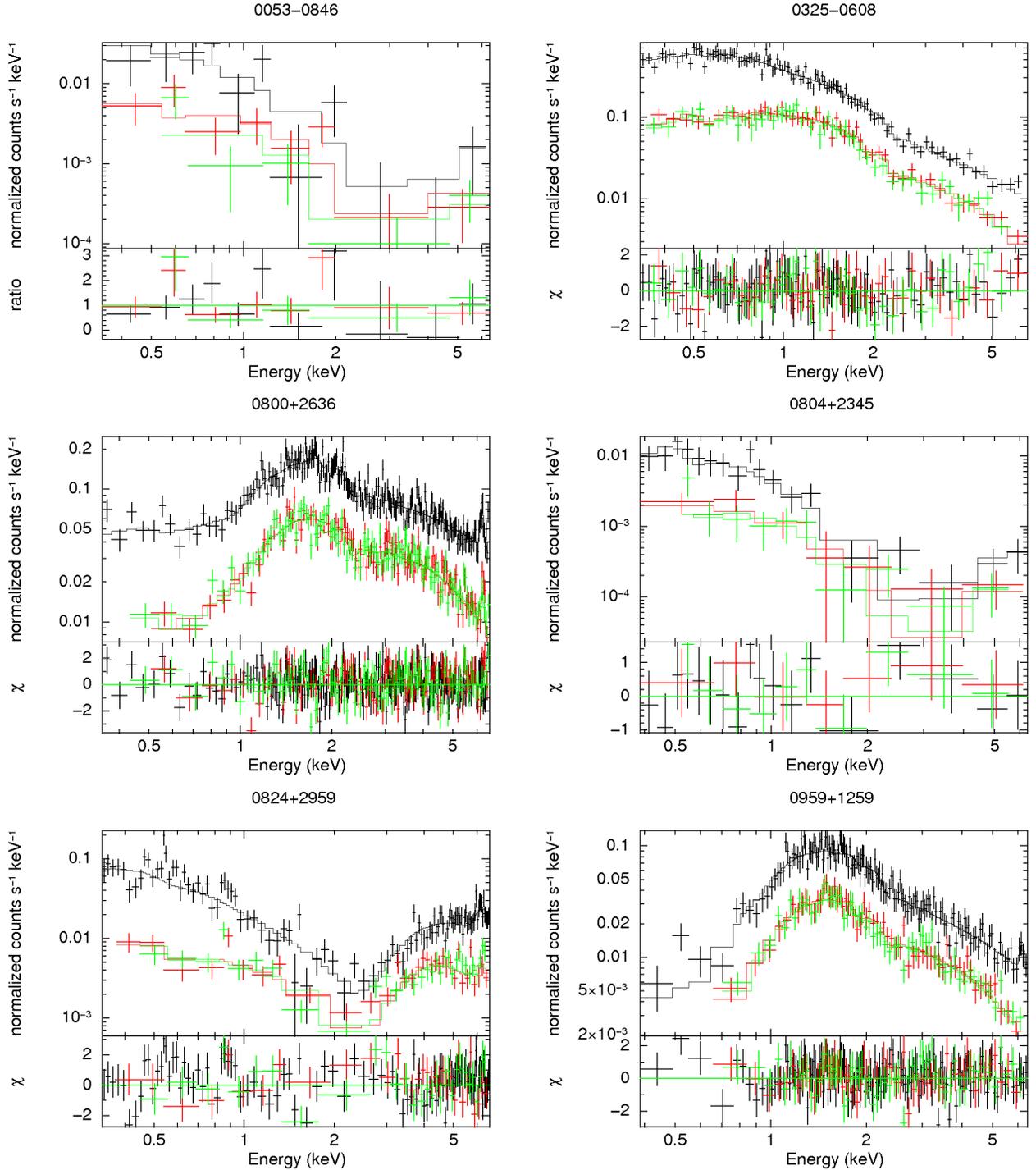}
\caption{\label{spec1}Best-fit spectra for the first 6 of our sources. The models are as follows: 0053-0846: double absorbed power law, 0325-0608: single absorbed power law + thermal component, 0800+2636: double absorbed power law + gaussian component to accommodate Fe K$\alpha$ line, 0804+2345: double absorbed power law, 0824+2959: double power law + gaussian component to accommodate Fe K$\alpha$ line, 0959+1259: single absorbed power law + gaussian component to accommodate Fe K$\alpha$ line. The black line indicates the PN spectra and the red and green lines denote the MOS1 and MOS2 spectra respectively.}
\end{figure}

\begin{figure}[f]
\plotone{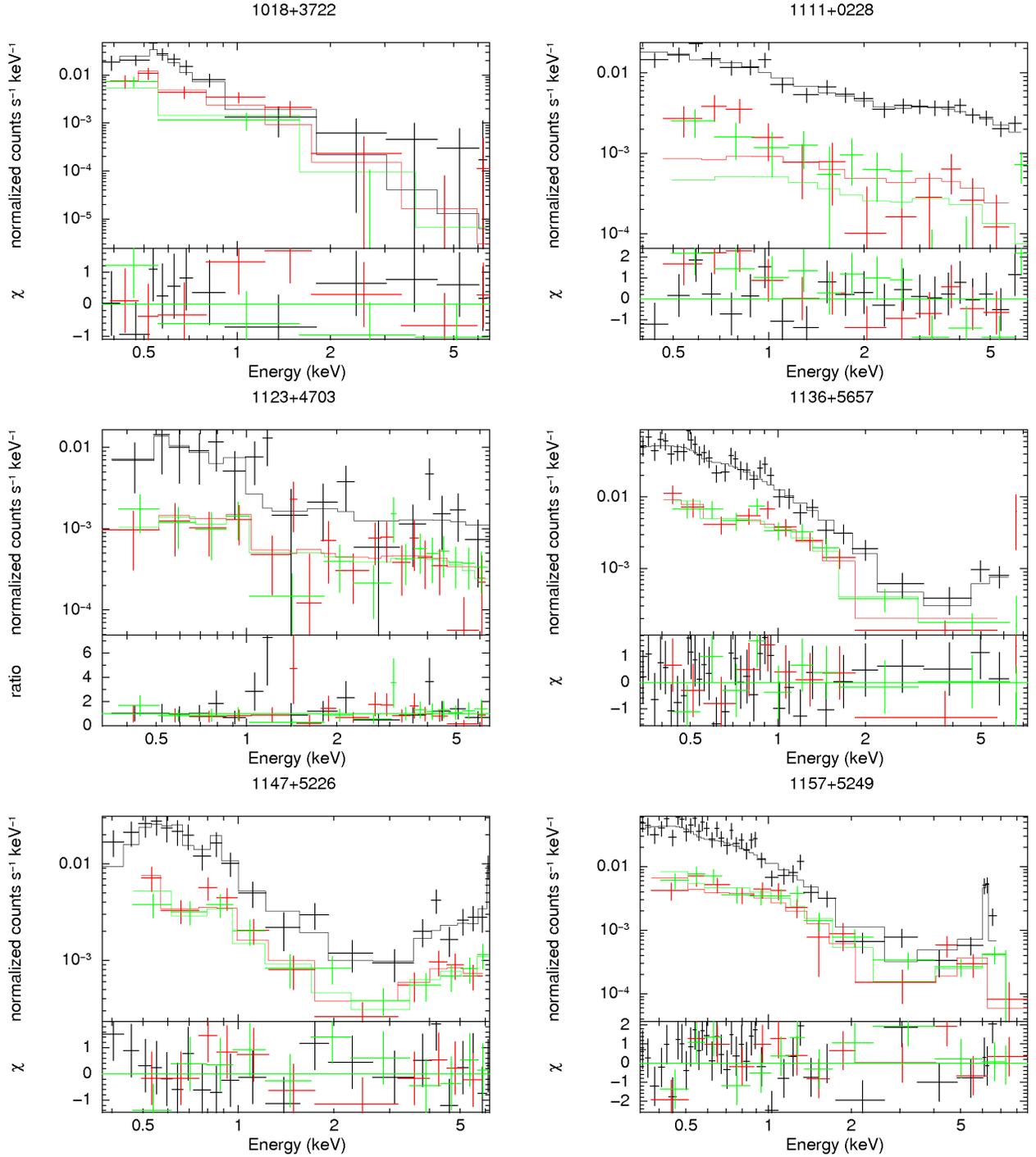}
\caption{\label{spec2}Best-fit spectra for the next 6 of our sources. The models are as follows: 1018+3722: single absorbed power law + thermal component, 1111+0228: double absorbed power law, 1123+4703: single absorbed power law + thermal component, 1136+5657: double absorbed power law, 1147+5226: double absorbed power law + thermal component + gaussian component to accommodate Fe K$\alpha$ line, 1157+5249: double absorbed power law + gaussian component to accommodate Fe K$\alpha$ line. The color coding is the same as Figure~1. }
\end{figure}

\begin{figure}[f]
\plotone{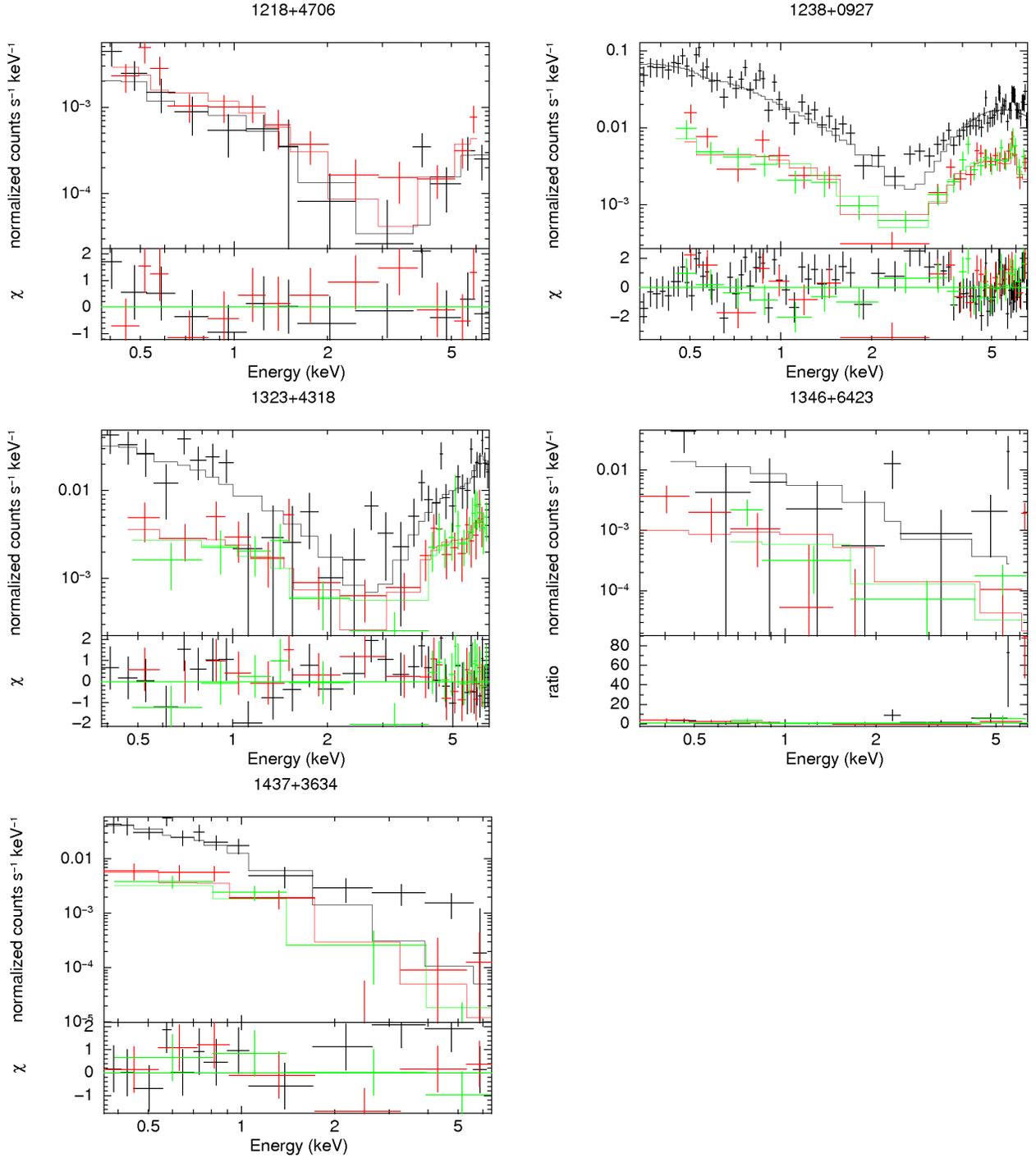}
\caption{\label{spec3}Best-fit spectra for the remainder of our sources. The models are as follows:  1218+4706: double absorbed power law, 1238+0927: double absorbed power law + gaussian component to accommodate Fe K$\alpha$ line, 1323+4318: double absorbed power law + gaussian component to accommodate Fe K$\alpha$ line, 1346+6423: single absorbed power law, 1437+3634: single absorbed power law. The color coding is the same as Figures~1 and 2, with the exception of 1218+4706 where the black and red lines indicate the MOS1 and MOS2 spectra respectively. }
\end{figure}

The spectra with the best-fit models are shown in Figures \ref{spec1}, \ref{spec2} and \ref{spec3}. The observed APEC and hard X-ray luminosities derived from the fits above are listed in Table~\ref{lumin}.\footnote{The hard X-ray luminosities for 0325-0608, 0804+2345, 1018+3722, 1123+4703, 1136+5657, 1147+5226 and 1157+5249 are based on the power-law component of the APEC + powerlaw fits as the addition of the APEC component improved the fit statistically and/or resulted in more physically reasonable fits for $\Gamma$ for these targets; the hard X-ray luminosities for the remaining targets are derived from the single/double power law fits. We note, however, that the differences between the luminosity derived using the APEC + power-law model vs. just the power-law model are minor.} The APEC luminosity in some cases may reflect the luminosity due to scattered soft X-ray flux rather than thermal emission from star formation processes.  Both the APEC and hard X-ray luminosities have not been corrected for absorption.

\section{Ancillary Data}

\subsection{Sloan Digital Sky Survey}

The methodology we used to derive the properties of the AGNs and their host galaxies are described in detail in K03, Heckman et al. (2004), and Kauffmann \& Heckman (2009). It is important to note that the SDSS spectra are obtained through fibers with angular diameters of 3 arcsec. This corresponds to typical projected dimensions of $\sim$ 2 to 3 kpc for our targets. This is large enough to encompass the AGN Narrow Line Region (e.g. Schmitt \& Kinney 1996; Schmitt et al. 2003). However, since the galaxies in our sample have typical optical diameters of 15 to 20 kpc (e.g. Kauffmann et al. 2003), the fiber only covers the inner region of the host galaxy (bulge and inner disk).

For the AGN we first compute an apparent [OIII]$\lambda$5007 luminosity using the observed emission-line flux. We then compute an extinction-corrected [OIII] luminosity using the measured flux ratio of the narrow H$\alpha$ and H$\beta$ emission lines and assuming an intrinsic ratio of 3.1 and the standard R = 3.16 extinction curve for galactic dust \citep{OF}. The median [OIII] extinction correction for our sample is 1.0 magnitude (with a range from 0.5 to 2.3 mag). 

To estimate a bolometric correction, we used the ratio of the mid-IR continuum luminosity from the Spitzer IRS data (see below) to the extinction-corrected [OIII] luminosities from SDSS for the present sample and then used the models of IR emission from obscuring tori in Type 2 AGN in Nenkova et al. (2008) to estimate the ratio of mid-IR and bolometric luminosity. The implied bolometric corrections to the extinction-corrected [OIII] luminosity are in the range of 500 to 900. In Heckman et al. (2004) we used the available multi-waveband data to show that for Type 1 Seyfert nuclei and QSOs, the average bolometric correction to the {\it uncorrected} [OIII] luminosity was 3500. Typical measured values for the Balmer decrement in Type 2 Seyferts imply extinction corrections of $\sim$0.8 to 2.0 magnitudes (e.g Kewley et al. 2006), so we would expect the typical bolometric correction to the extinction corrected [OIII] luminosity for such AGN to be in the range $\sim$ 550 to 1700. This range is consistent with what we infer for our sample based on the mid-IR data. We adopt a bolometric correction of 700 for our sample.

We use the observed correlation between black hole mass $M_{BH}$ and bulge velocity dispersion $\sigma_*$ \citep{Tremaine} to derive an estimated black hole mass for each galaxy. We then use the ratio of the extinction-corrected [OIII]$\lambda$5007 emission-line luminosity and the black hole mass as a proxy for the Eddington ratio ($L_{AGN}$/L$_{Edd}$). For reference, a ratio of $L_{[OIII]}/M_{BH} \sim$50 corresponds roughly to $L_{AGN}/L_{Edd}$ = 1 (Heckman et al. 2004; Kauffmann \& Heckman 2009).

For ordinary star-forming galaxies, star formation rates can be derived for the region sampled by the spectroscopic fiber by using the extinction-corrected luminosity of the H$\alpha$ emission-line, together with the additional information about nebular conditions provided by the [OIII]5007, [NII]6584, and [SII]6717,6731 lines (Brinchmann et al. 2004). However, in our objects these lines will be strongly contaminated by the contribution of the AGN. Since we can not directly calculate star formation rates using these nebular emission-lines, we instead use the amplitude of the 4000\AA\ break (D$_n$(4000)), and the strong inverse correlation between $D_n(4000)$ and specific star-formation rate ($SFR/M_*$) found for galaxies without AGN (Brinchmann et al. 2004). Brinchmann et al. show that typical uncertainties in log($SFR/M_*$) estimated this way are about $\pm$ 0.5 dex (one $\sigma$) for any individual object. We have verified these estimates using the positive correlation between the strength of the high-order stellar Balmer absorption-lines and $SFR/M_*$ \citep{Chen}. In one object (1218+4706) the small value for both D$_n(4000)$ and the equivalent widths of the Balmer absorption-lines imply that the spectrum is significantly contaminated by scattered continuum from the hidden Type 1 AGN. We are thus not able to estimate a star formation rate for this galaxy from the SDSS spectrum.

The parameters derived from the SDSS spectra are listed in Table \ref{spitzer}.

\subsection{Spitzer Space Telescope}

The data used in this paper are taken from the analysis reported in Heckman et al. (2009). In brief, the {\it Infrared Spectrograph (IRS)} on the {\it Spitzer Space Telescope} was used to obtain low-resolution ($64 < R < 128$) spectra with the Short-Low (3.7'' $\times$ 57'' aperture) and Long-Low modules (10.7'' $\times$ 168'' aperture) and high-resolution (R $\sim$ 650) spectra with the Short-High (4.7'' $\times$ 11.3'' aperture) and Long-High modules (11.1'' $\times$ 22.3'' aperture). Standard observing and data reduction procedures were followed (see Heckman et al. 2009 for details).

For the purposes of this paper, we will utilize two diagnostics of the luminosity of the AGN. The first is the luminosity of the [OIV] 25.89$\mu$m emission-line. Since the production of OIV requires photons with energies above 54.9 eV (above the He II edge at 54.4 eV) it is weak in purely star forming galaxies. Thus, it is a good tracer of the hard AGN ionizing continuum (Mel\'{e}ndez et al. 2008). Since it is produced in the Narrow-Line Region, it should suffer minimal obscuration in our AGN. We use [OIV] as opposed to the higher ionization [Ne  V] 14.32,24.32 $\mu$m lines in the mid-infrared as the [OIV] line is stronger (has larger signal-to-noise ratio).

The second diagnostic is the luminosity of the mid-infrared continuum. This is produced by the warm dust in the obscuring torus. As described in Heckman et al. (2009) we have defined two windows in the mid-infrared spectrum that are devoid of significant emission or absorption features: one is centered at 13.5$\mu$m and the other at 30$\mu$m. The mid-infrared continuum luminosity we use in this paper is the sum of the monochromatic luminosities ($\nu L_{\nu}$) at the midpoints of these two bands. In principle, dust emission associated with a starburst can also contribute to the mid-IR continuum. To assess this, we have used the correlations between the mid-IR continuum luminosity and the luminosities of the 6.2, 7.7, 11.3, and 17.3 $\mu$m PAH features seen in starbursts (Smith et al. 2007). Using the PAH luminosities in our Type 2 AGN, we then find that the estimated starburst contribution to the mid-IR continuum is typically only about ten percent, with a range from a few percent to a few tens of percent. We will discuss this in detail in a future paper in this series.

The parameters derived from the {\it Spitzer} data are listed in Table \ref{spitzer}.

\section{Results}

\subsection{The Relative Strength of the Hard X-ray Continuum: Comparison with Previous Samples}

As discussed in the Introduction, it is well known that the hard X-ray emission is very weak in many obscured AGN. This can be quantified by comparing the ratio of the hard X-ray luminosity to that of proxies for the bolometric luminosity of the AGN. In this section we will consider four such proxies: 1) the observed luminosity of the [OIII]$\lambda$5007 emission-line (L$_{[OIII],obs}$, as used in e.g. Vignali et al. 2006, Ptak et al. 2006) 2) the luminosity of the [OIII]$\lambda$5007 line, corrected for dust extinction (L$_{[OIII],corr}$), listed in Table \ref{lumin} 3) the luminosity of the [OIV] 25.89$\mu$m line (L$_{[OIV]}$) and 4) the luminosity of the mid-infrared (13.5 plus 30$\mu$m) continuum (L$_{MIR}$). In all cases, the hard X-ray luminosity is measured in the 2-10 keV band (L$_{2-10keV}$) based on the fits discussed in section 2 above.

\begin{figure}[f]
\epsscale{1.2}
\plottwo{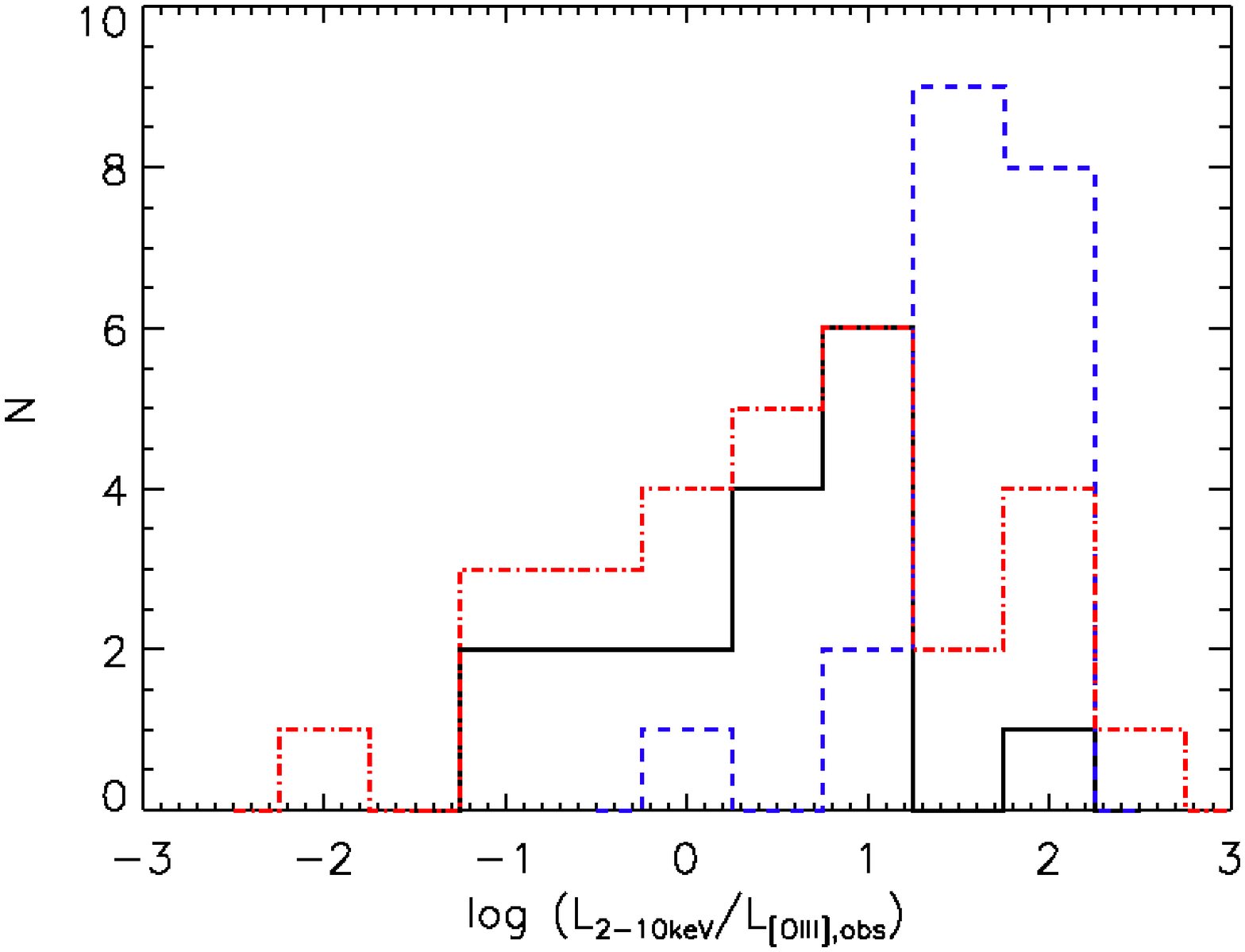}{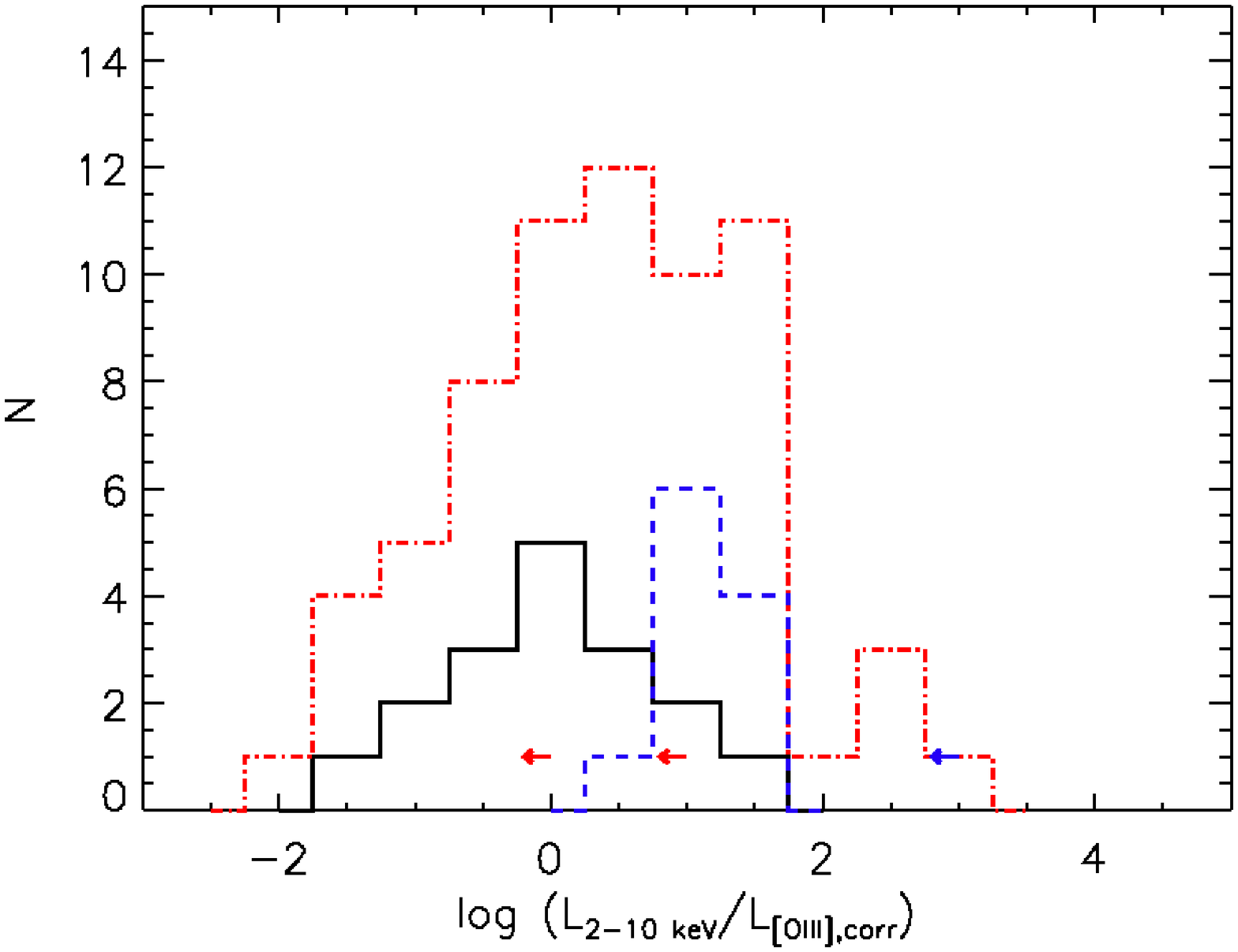}
\caption{\label{hist}Left: Histogram of L$_{2-10 keV}$/L$_{[OIII],obs}$ values. Our sample is represented by the solid black line, the Heckman et al. 2005 Sy1 sample is depicted by the dashed blue line, and the Heckman et al. 2005 Sy2 sample is shown by the dotted-dashed red line. Right: Histogram of L$_{2-10 keV}$/L$_{[OIII],corr}$ values. Our sample is the solid black line. The dashed blue line represents the Sy1s from the Mulchaey et al. 1994 sample (where $L_{[OIII]}$ is corrected for extinction using the Balmer decrement) and the dotted-dashed red line denotes the Sy2s from the Bassani et al. 1999 sample. The blue (red) arrow(s) represent an upper L$_{2-10 keV}$/L$_{[OIII],corr}$ limit on the Mulchaey et al. (Bassani et al.) sample. In both figures, our sample of Type 2 AGN shows a very broad range in the luminosity ratio and the distribution is displaced well below that of the Type 1 AGN. This implies significant X-ray absorption in the Type 2 AGN.}
\end{figure}

We begin with the left panel of Figure \ref{hist} with a plot of the distribution of the ratio of L$_{2-10keV}/$L$_{[OIII],obs}$ for our sample. There is a huge range in the luminosity ratio (roughly four orders-of-magnitude). In order to assess the effect of X-ray absorption, we compare this distribution to that for a sample of ``[OIII]-bright'' Type 1 (dashed blue line) AGN taken from Heckman et al. (2005). This comparison shows that that distribution of hard X-ray luminosities for our sample of Type 2 AGN is significantly displaced below the peak of the narrow distribution of the Type 1 AGN. For completeness we also show the distribution of L$_{2-10keV}/$L$_{[OIII],obs}$  for the sample of Type 2 AGN (dot-dashed red line) from Heckman et al. (2005). These objects were drawn from many different heterogeneous samples of AGN rather than representing a complete and homogeneously selected sample like ours. Nevertheless, the distributions in the two samples of Type 2 AGN roughly agree and result in a Kolmogorov-Smirnov P-value of 0.07, i.e., the distributions differ at a significance of only $\sim$1.5$\sigma$ (note that the K-S test does not incorporate errors in the data points of the samples being compared). We also compare our results to X-ray observations of higher redshift (0.3 $< z <$ 0.8) Type II AGN selected based on [OIII] luminous quasars from SDSS \citep{Z03}. Ptak et al. (2006) and Vignali et al. (2006) find that a high fraction of their sources likely suffer from significant absorption: 6 out of 8 (Ptak et al. 2006, based on spectral fits for 5 sources and the L$_{2-10 keV}$/L$_{[OIII],obs}$ ratio for 1 source) and 10 out of 16 (Vignali et al. 2006, based on spectral fits for 2 sources and the L$_{2-10 keV}$/L$_{[OIII],obs}$ ratio for 8 sources).

The [OIII]$\lambda$5007 line will be affected by dust obscuration associated with the Narrow Line Region and the interstellar medium of its host galaxy. Thus, a better indicator of the intrinsic AGN luminosity is provided by the extinction-corrected [OIII]$\lambda$5007 luminosity (e.g. Lamastra et al. 2009). In the right panel of Figure \ref{hist} we show the distribution of L$_{2-10keV}/$L$_{[OIII],corr}$ for our sample. For comparison we show the distribution for Type 1 AGN taken from Mulchaey et al. (1994), corrected for extinction using their published Balmer decrement. These sources are represented by the dashed blue line in the right panel of Figure \ref{hist} and blue left arrow. Again, we find that the Type 2 AGN in our complete sample span a very large range in the relative strength of the hard X-rays and are significantly weaker hard X-ray emitters than the Type 1 AGN. All but one of our AGN have luminosity ratios lying below the median value for the Type 1 AGN (L$_{2-10keV}/$L$_{[OIII],corr} \sim$ 10), and 11 of the 17 lie about an order-of-magnitude or more below this value, the nominal Compton-thick boundary (L$_{2-10keV}/$L$_{[OIII],corr} \leq$ 1, e.g. Bassani et al. 1999, where Compton-thick sources, classified as such by N$_H$ and/or Fe K$\alpha$ EW measurements, mostly fall below this boundary). Our results agree with those found for a heterogeneous sample of Type 2 AGN by Bassani et al. (1999), denoted by the dotted-dashed red line and red left arrows in Figure \ref{hist}. For our sample we find a mean log (L$_{2-10keV}$/L$_{[OIII],corr}$) value of 0.003 with dispersion 0.75 dex, compared with 1.09 $\pm$ 0.63 dex for a heterogeneous sample of 61 predominantly Compton-thin Type 2 AGN (Lamastra et al. 2009). Our lower $L_{2-10keV}$/L$_{[OIII],corr}$ mean value imply that our targets are more heavily obscured while the larger dispersion indicates the wide range of L$_{2-10keV}$/L$_{[OIII],corr}$ values present in our sample.

\begin{figure}[f]
\epsscale{1.2}
\plottwo{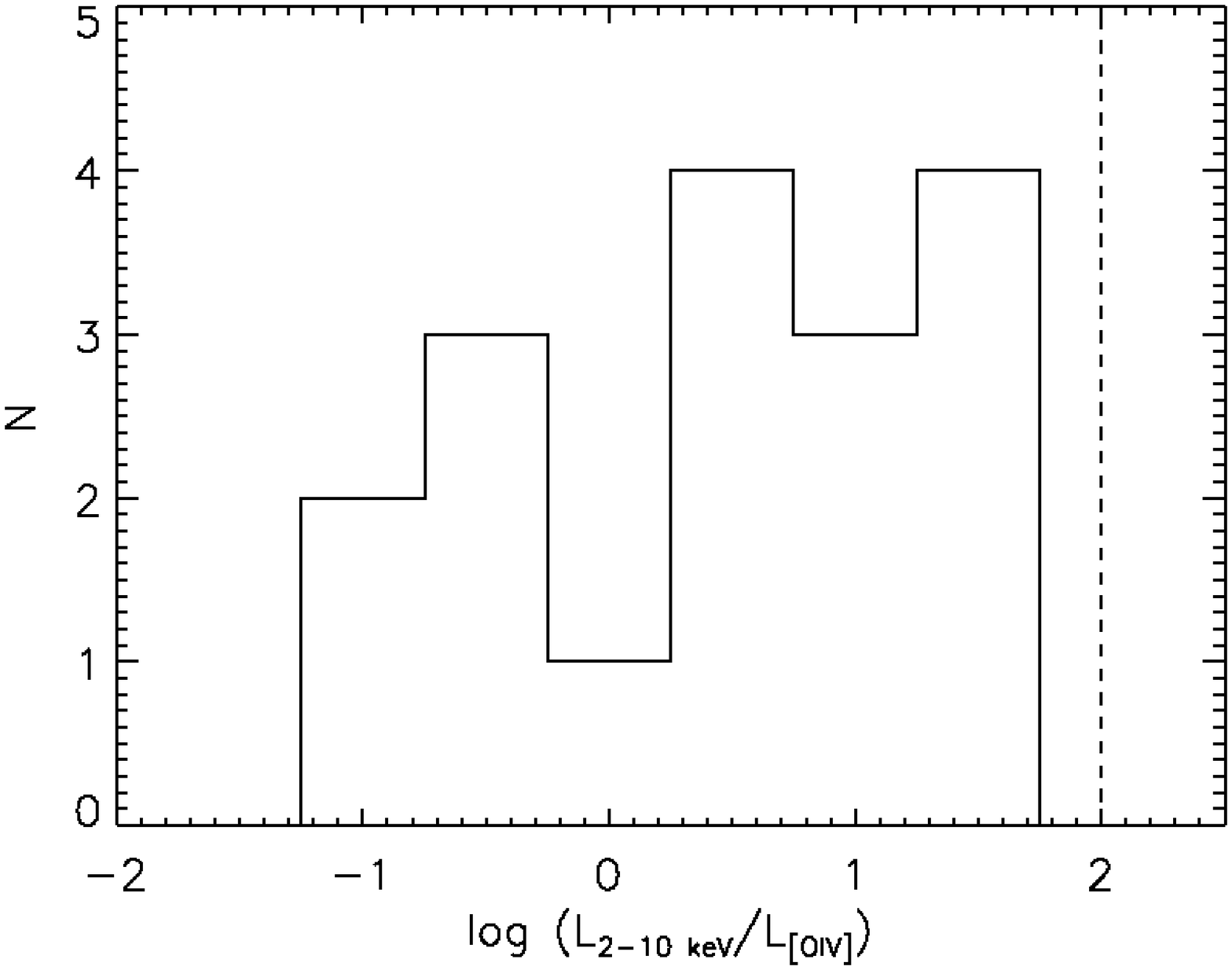}{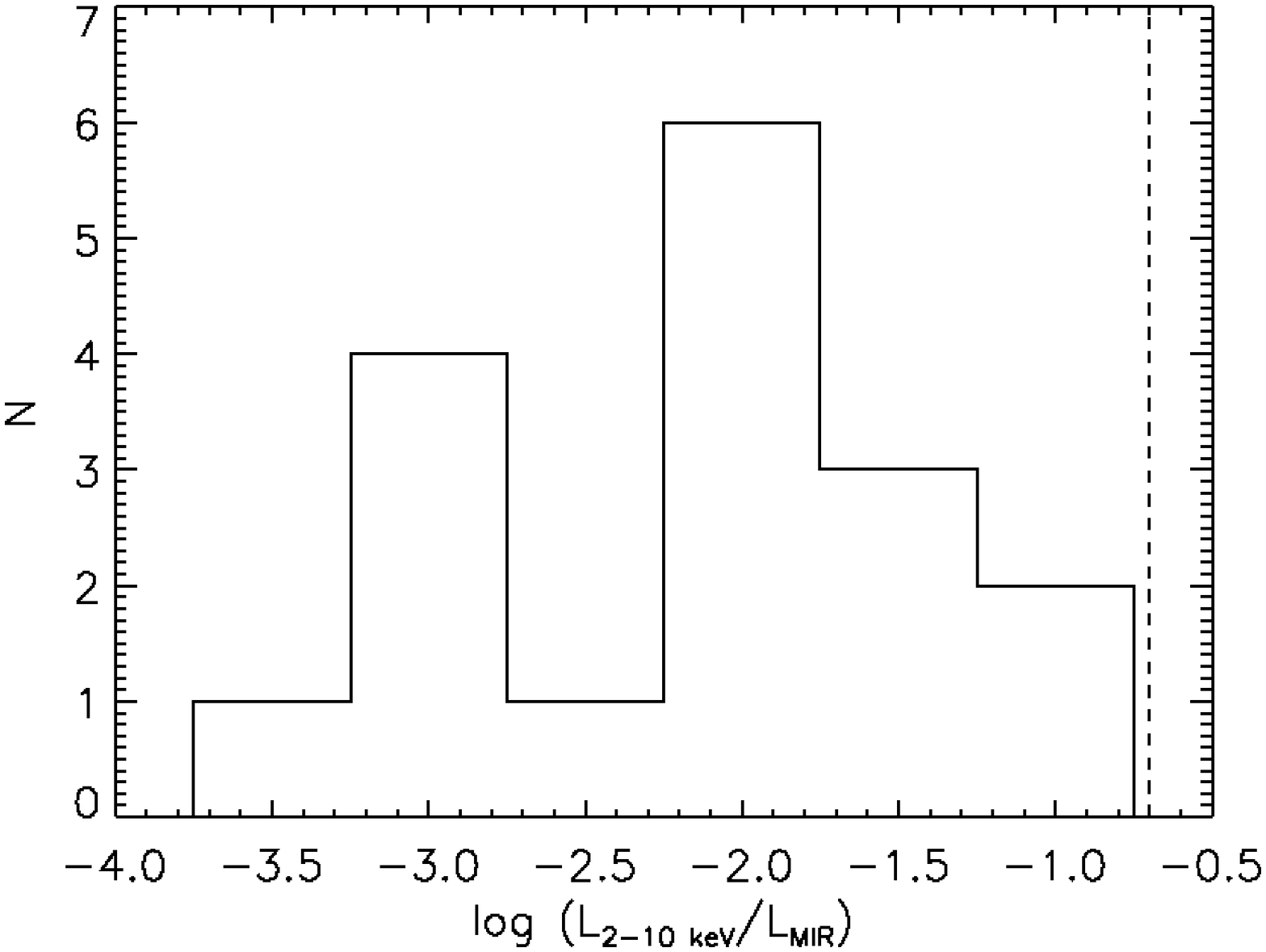}
\caption{\label{hist_oiv}Left: Histogram of L$_{2-10 keV}$/L$_{[OIV]}$ values for our sample of Type 2 Seyferts. The dashed vertical line indicates the mean value for Type 1 Seyferts \citep{OIV}. Right: Histogram of L$_{2-10 keV}$/L$_{MIR}$ values. The dashed vertical line indicates the mean value for Type 1 Seyferts \citep{Marconi}. The distribution in the luminosity ratio in our sample is very broad in both figures and extends nearly four orders-of-magnitude below the values seen in Type 1 AGN. This implies significant X-ray absorption in Type 2 AGN.}
\end{figure}

Finally, we show the corresponding distributions for the ratio of the hard X-ray luminosity to the luminosity of the [OIV] 25.89$\mu$m line (Figure \ref{hist_oiv}, left) and the mid-IR continuum (Figure \ref{hist_oiv}, right). These measures should be unaffected by obscuration, and confirm the wide spread in the relative strength of the hard X-ray emission in Type 2 AGN. In both plots, all the Type 2 AGN lie below the mean values for Type 1 AGN: log (L$_{2-10keV}/$L$_{[OIV]}$) $\sim$ 2.1 with a dispersion of 0.3 dex (Mel\'{e}ndez et al. 2008) and log (L$_{2-10keV}$/L$_{MIR}$) $\sim -0.70$ with a dispersion of 0.25 dex (Marconi et al. 2004 \& Elvis et al. 1994). The majority (10-12/17) lie more than a factor of ten below these values.

The significance of our results is that we have been able to quantify the distribution of the relative strength of the hard X-rays for the first time using a complete and homogeneous sample of optically selected Type 2 (obscured) AGN, and using both optical and mid-infrared indicators of the intrinsic AGN luminosity. All 3 methods identify the same 11 targets as suffering from heavy obscuration: 0053-0846, 0804+2345, 1018+3722, 1111+0228, 1123+4703, 1136+5656, 1147+5226, 1157+5249, 1218+4705, 1346+6423 and 1437+3634.

\subsection{Clues from the Fe K$\alpha$ Line}

One immediate question is whether the wide range in the relative strength of the hard X-rays seen above is mostly due to obscuration, or whether it could indicate that the hard X-ray emission is intrinsically weak in Type 2 AGN (in violation of the standard unified model). Several such cases of ``unabsorbed'' Type 2 AGN have been proposed (e.g. Panessa \& Bassani 2002; Brightman \& Nandra 2008). One way to discriminate between these possibilities is to use the Fe K$\alpha$ emission line to look for evidence of a heavily obscured AGN (e.g. Bassani et al. 1999; Levenson et al. 2006). In such cases, the equivalent width of the K$\alpha$ line can be very large ($\sim$ a keV) because the X-ray continuum at 6.4 keV traverses a significantly larger gas column than the K$\alpha$ line emission, and is therefore more much highly obscured (e.g. Krolik \& Kallman 1987; Levenson et al. 2006; Murphy 2008).

\begin{figure}[f]
\epsscale{1.2}
\plottwo{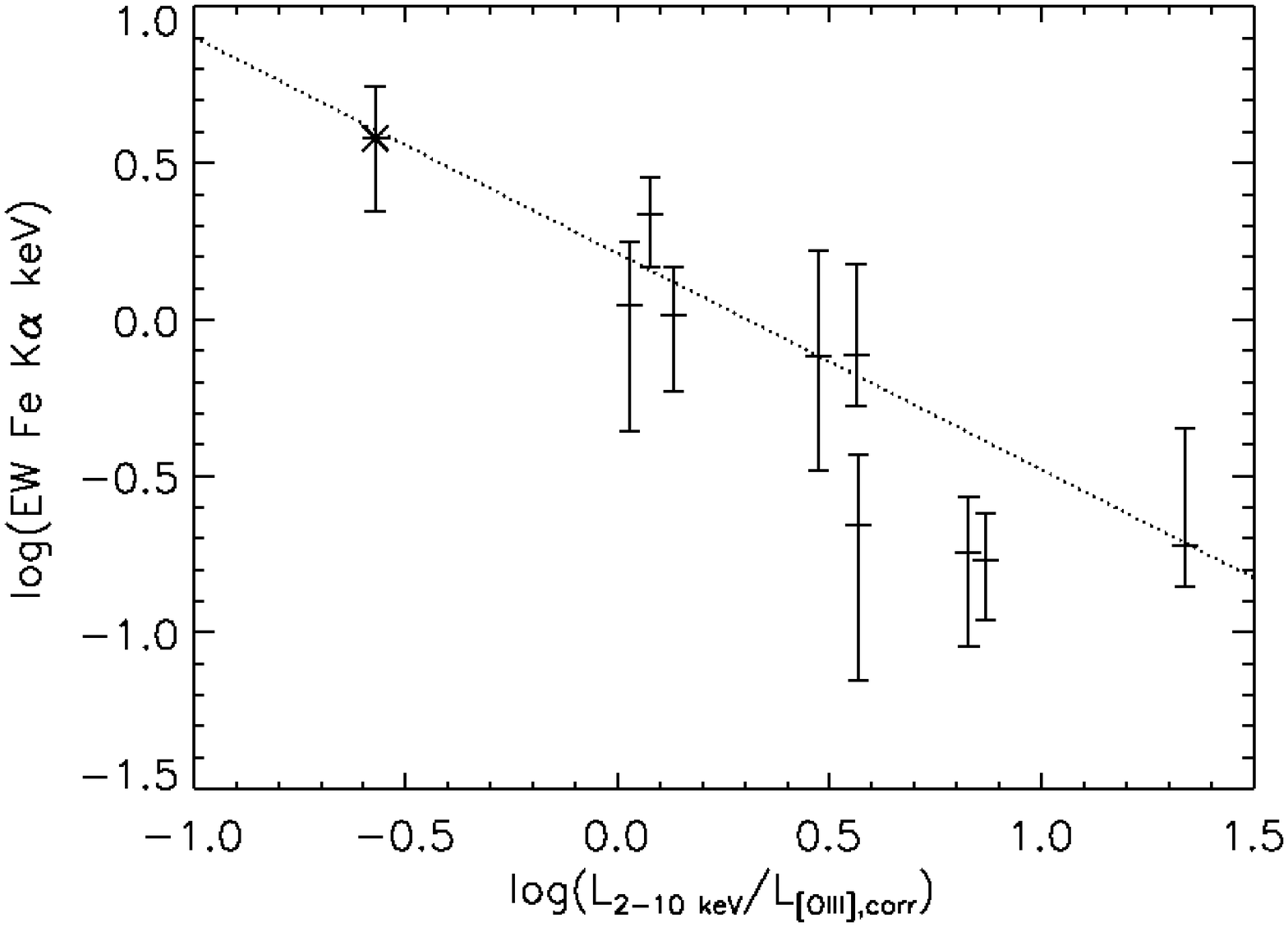}{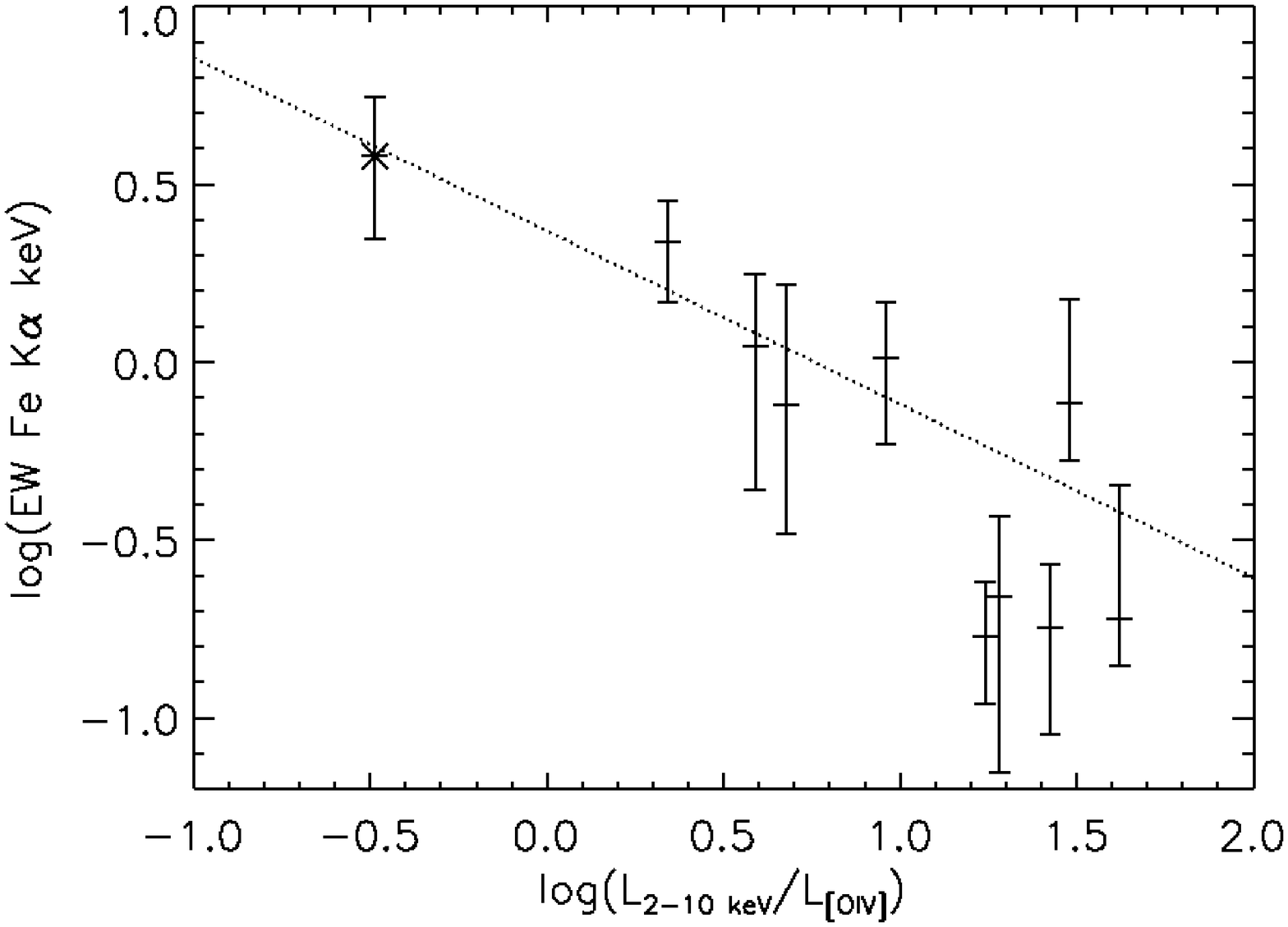}
\caption{\label{ews}Left: L$_{2-10keV}$/L$_{[OIII]}$ vs. Fe K$\alpha$ EW with best-fit correlation overplotted. Right: L$_{2-10keV}$/L$_{[OIV]}$ vs. Fe K$\alpha$ EW with best-fit correlation overplotted. In both plots, the asterisk represents the coadded data for the Compton-thick candidates. Both relations are significant at greater than the 99.8\% confidence level. The inverse correlations are consistent expectations for a heavily obscured hard X-ray source.}
\end{figure}

As shown in Bassani et al. (1999), this picture is supported by an inverse correlation observed between the equivalent width of the K$\alpha$ line (EW) and the ratio of hard X-ray and [OIII]$\lambda$5007 emission-line luminosities. We therefore investigate the relationships between the Fe K$\alpha$ EW and the luminosity ratios described above. For this analysis, we coadded the MOS spectra for the eight Compton-thick candidates (L$_{2-10keV}$/L$_{[OIII],corr} \leq 1$) that had unconstrained Fe K$\alpha$ EW or upper limits on the Fe K$\alpha$ line flux (0053-0846, 0804+2345, 1018+3722, 1111+0228, 1123+4703, 1136+5657, 1346+6423 and 1437+3634); these targets are flagged in Table \ref{Fe}. After stacking these spectra with the \textit{ftools} routine \textit{addspec}\footnote{All the sources are on-axis and at relatively low redshift, so the spectra can be simply added}, we find a coadded total EW value of $\sim$ 3.8$^{+1.8}_{-1.5}$ keV and a coadded L$_{2-10keV}$ value of $\sim6\times 10^{40}$ erg s$^{-1}$, compared with an EW value of 1.96$^{+0.90}_{-0.79}$ and  L$_{2-10keV}$ value of $\sim1\times 10^{41}$ erg s$^{-1}$ from simultaneously fitting the coadded MOS spectra for each of these sources. We performed a linear regression fit on both L$_{2-10keV}/$L$_{[OIII],corr}$ and L$_{2-10keV}/$L$_{[OIV]}$ {\it vs.} Fe K$\alpha$ EW using these coadded data for the 8 Compton-thick cases and Fe K$\alpha$ EW detections. The data and resulting fits are shown in Figure \ref{ews}. We find a slope of -0.69 with a dispersion $\sigma$=0.23 dex for L$_{2-10keV}/$L$_{[OIII],corr}$ (compared to a slope of -0.37 with $\sigma$=0.05 dex from the Bassani et al. 1999 sample\footnote{This relationship applies to a subset of 52 Seyfert 2 galaxies from the Bassani et al. 1999 sample that have extinction corrected $f_{[OIII]}$ values and Fe K$\alpha$ measurements.}) and a slope of -0.49 and $\sigma$=0.16 dex for L$_{2-10keV}/$L$_{[OIV]}$. Both correlations are significant at greater than the 99.8\% confidence level. These results, a statistically significant anti-correlation between L$_{[2-10keV]}$/L$_{[OIII]}$ and Fe K$\alpha$ EW, imply that the weak hard X-ray emission in our sample is due to obscuration, rather than intrinsically weak X-ray emission.

\begin{figure}[f]
\epsscale{1.2}
\plottwo{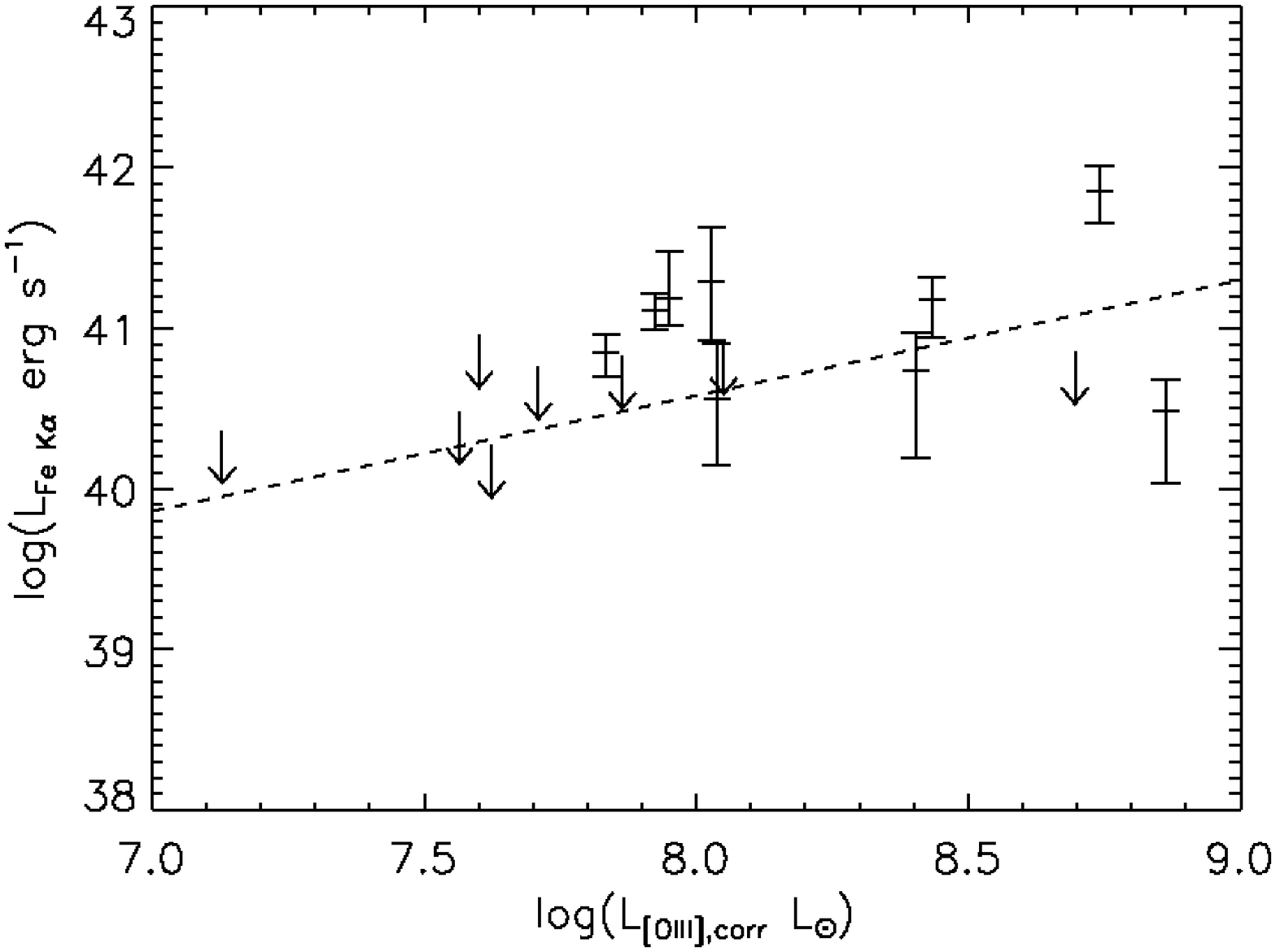}{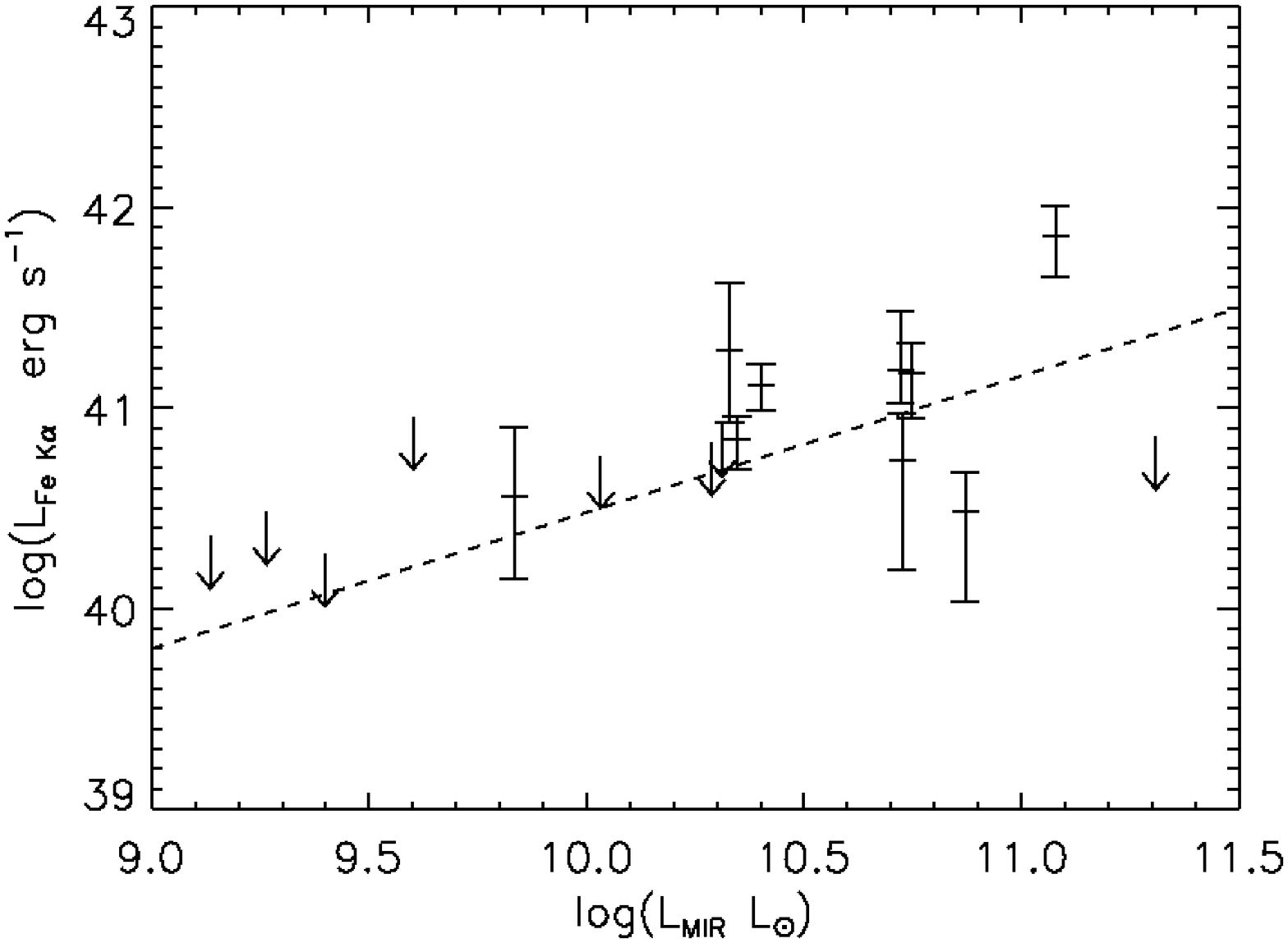}
\caption{\label{alpha_oiii}Left: Extinction corrected L$_{[OIII]}$ vs. Fe K$\alpha$ luminosity with best-fit correlation from survival analysis overplotted. Right: L$_{MIR}$ vs. Fe K$\alpha$ luminosity with best-fit correlation from survival analysis overplotted. The L$_{[OIII]}$-Fe K$\alpha$ correlation is significant at about the 94\% confidence level and the L$_{[MIR]}$-Fe K$\alpha$ correlation is significant at about the 99\% confidence level. The inverse correlations are consistent with those expected in the cases of heavily obscured AGN.}
\end{figure}

Ptak et al. (2003) suggested that the luminosity of the Fe K$\alpha$ line itself could be used as an estimator of the AGN bolometric luminosity in obscured AGN: they found a correlation between extinction-corrected [OIII] luminosity and Fe K$\alpha$ luminosity for both Seyfert 1 and 2 galaxies (slope of 1 with a scatter of $\pm$ 0.5 dex). As shown in Figure \ref{alpha_oiii}, we do see significant correlation between L$_{[OIII],corr}$ and L$_{Fe K\alpha}$. Using survival analysis to calculate the correlation coefficient between these two parameters (\textbf{ASURV} Rev 1.2 Isobe \& Feigelson 1990; LaValley, Isobe, \& Feigelson 1992) for bivariate data (Isobe, Feigelson, \& Nelson 1986), we find a slope of 0.7 $\pm 0.3$ with a scatter of $\sigma$=0.5 dex and a correlation significant at the $\sim94\%$ confidence level. When comparing L$_{MIR}$ with L$_{Fe K\alpha}$, we find a similar correlation: a slope of 0.7 $\pm$ 0.2, $\sigma$ of 0.4 dex, and a $\sim99\%$ correlation probability. However, the correlation between L$_{[OIV]}$ and L$_{Fe K\alpha}$ is only significant at the $\sim85\%$ confidence level.

\subsection{Estimation of the Absorbing Column Density}

\begin{figure}[f]
\epsscale{0.60}
\plotone{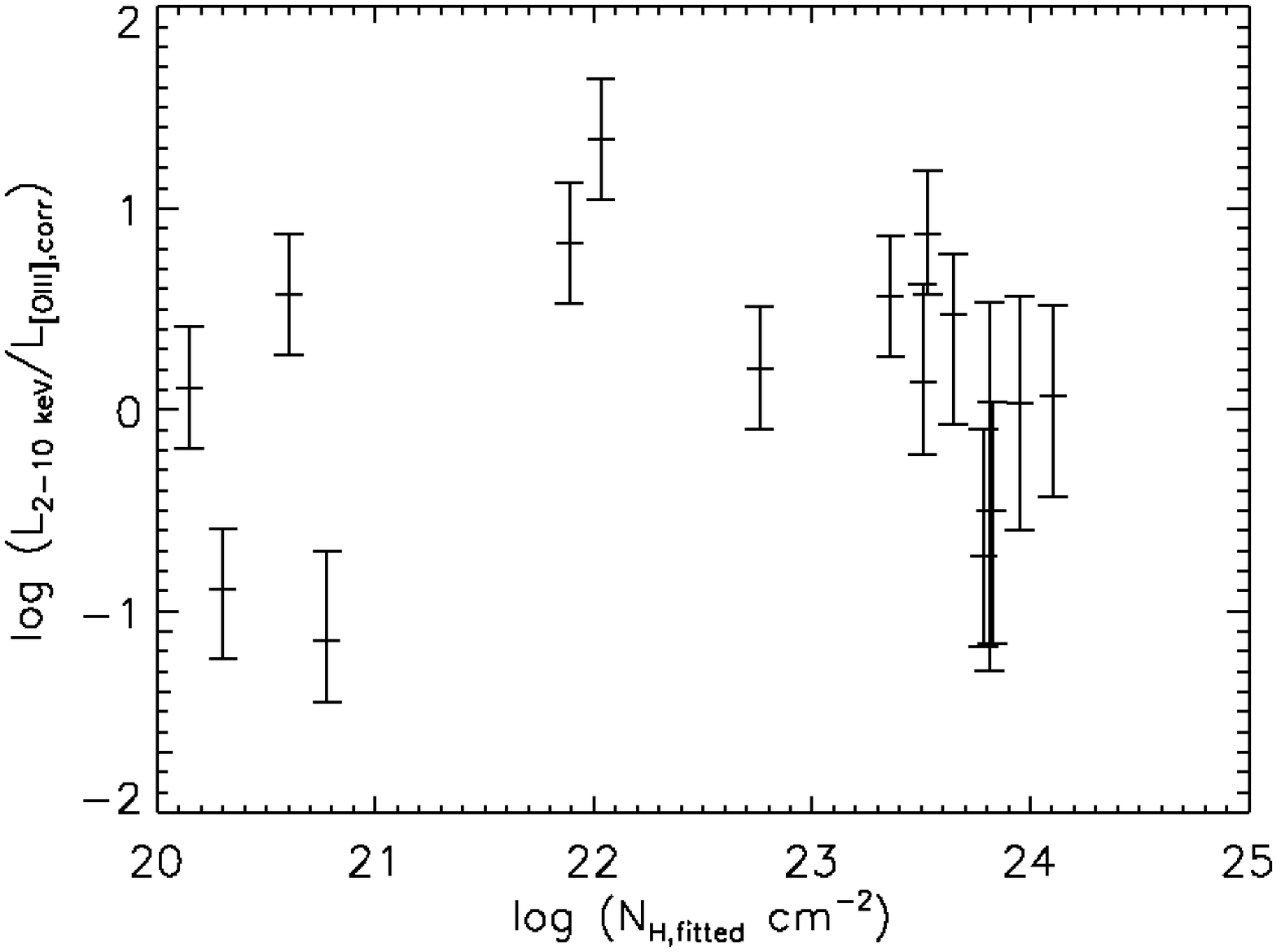}
\caption{\label{nh_fit_ratio}L$_{2-10keV}$/L$_{[OIII],corr}$ vs. N$_{H,fitted}$. The lack of any correlation implies that the fitted column densities do not represent the actual values responsible for the X-ray absorption.}
\end{figure}

\begin{figure}[f]
\epsscale{0.60}
\plotone{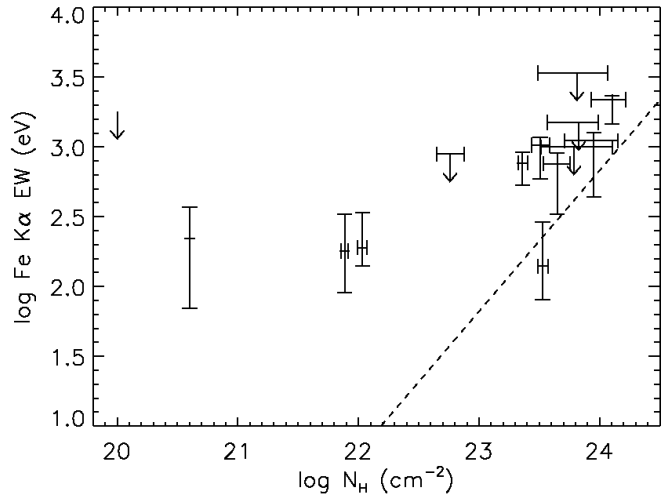}
\caption{\label{ew_nh}The fitted column density $N_{H,fitted}$ vs. Fe K-$\alpha$ EW. The dashed line represents the relationship from Krolik \& Kallman (1987) for Fe K$\alpha$ EW as a function of $N_H$ for an AGN oriented face-on. Our Fe K$\alpha$ EW values are systematically higher than this relation, implying that the Fe K$\alpha$ photons are not produced in regions consistent with transmission along the line of sight with the fitted N$_H$ values. Additional absorption, from matter out of the line of sight, may be required to accommodate the Fe K$\alpha$ EW values. Unlike Figure \ref{ews}, we plot the individual EW upper limits rather than the coadded value for the Compton-thick candidates without Fe K$\alpha$ detections.}
\end{figure}

The results above imply that the hard X-ray emission typically suffers significant attenuation. For a simple geometry of a foreground slab of absorbing gas, we would expect that the absorbing column densities we derive from our spectral fits would be consistent with the attenuation inferred from the relative strength of the X-ray continuum (Figures \ref{hist} and \ref{hist_oiv}). This is clearly not always the case, as we show in Figure \ref{nh_fit_ratio} which plots the fitted column density N$_{H,fitted}$ {\it vs.} log (L$_{2-10keV}/$L$_{[OIII],corr}$). No correlation is present, and there are several noteworthy cases of apparently Compton-thick AGN (on the basis of a low L$_{2-10 keV}$/L$_{[OIII],corr}$ ratio) with small values of N$_{H,fitted}$. A similar result can be seen in Figure \ref{ew_nh} where we plot the EW of the Fe K$\alpha$ line vs. N$_{H,fitted}$. Krolik \& Kallman (1987) predicted the observable Fe K$\alpha$ equivalent width (EW) as a function of column density for the cases where the the AGN is oriented face-on and where the torus completely blocks the central engine. Using the parameters from Ptak et al. 1996 ($\Gamma$=1.8, $N_H \leq 10^{24}$ cm$^{-2}$ and face-on orientation), we plot this relationship in Figure \ref{ew_nh}. The EW values are systematically higher than those predicted by Krolik \& Kallman, and there are several cases with low apparent column densities yet strong K$\alpha$ lines. This implies that the Fe K$\alpha$ emission is produced in regions that are not consistent with transmission along the line of sight. Taken together, Figures \ref{nh_fit_ratio} and \ref{ew_nh} imply that the simple spectral models we have used do not adequately recover the true absorbing column densities in many cases as the transmitted component below 10 keV is obscured and only the reflected and/or scattered light is observed.

We have therefore taken a more complex (but more physically realistic) model for the absorption, based on using the observed relative strength of hard X-rays as a constraint: we compared the observed L$_{2-10keV}$ value with an expected X-ray luminosity in the absence of absorption and attributed the diminution between these two values to obscuration. Specifically, we have drawn 1000 random values of L$_{2-10 keV}$/L$_{[OIII],obs}$ from a Gaussian distribution with the same mean and dispersion as the Heckman et al. (2005) Type 1 Seyfert sample. By multiplying these values by the L$_{[OIII],obs}$ for each target in our sample, we derived a simulated unobscured L$_{2-10 keV}$ for each source. We then compared this unobscured L$_{2-10 keV}$ value with the observed L$_{2-10 keV}$, attributing the luminosity diminution to absorption. This approach is similar to the one used by Mel\'{e}ndez et al. (2008), who used the ratio of the hard X-rays to the [OIV] emission-line.

We ran this simulation using an absorbed partial covering model, where a certain fraction of the X-ray emission is obscured and the remaining emission is scattered into the observer's line of sight (i.e. a double absorbed power law model with a fixed ratio between the normalizations of the two power law components and the same $\Gamma$). Such a model may better emulate the physical properties of the obscuring torus than using a foreground slab of absorbing gas (i.e. a single absorbed power law model). In our simulations, we employed a range of input parameters, with the covering fraction ranging from 0.99 to 0.96 and $\Gamma$ ranging from 1.6 to 2.0 (from the CDF-S survey, $< \Gamma_{AGN} > =$ 1.75 $\pm$ 0.02 with $\sigma =$ 0.3 for a sample of 82 Type 1 and 2 X-ray bright ($>$180 net counts, or $>$80 counts in the 2-10 keV range) AGN, Tozzi et al. 2006). For each set of input parameters, the column density was varied from 0, corresponding to an unobscured source, to 10$^{25}$ cm$^{-2}$, with the count rate computed at each step. The simulated unobscured luminosity was normalized by the model luminosity at $N_H$=0. We then found the $N_{H,sim}$ that corresponds to the observed count rate by computing the model predicted count rate (i.e. the ratio of the observed count rate to the predicted unabsorbed count rate multiplied by the count rate for the $N_{H,sim}$=0 case) and interpolating the effective $N_{H,sim}$ that would produce this model count rate. The results are summarized in Table \ref{gamma}, for simulations where $\Gamma$ was held at 1.8 and the covering fraction was varied, and Table \ref{cf}, for simulations where the covering fraction is held at 0.99 and $\Gamma$ was varied. This range of covering fraction and $\Gamma$ was chosen to assess how the implied absorption varies with the covering fraction and power law index (i.e. the is range is consistent with many Seyfert 2 galaxies but does not span the full parameter for absorbed AGN). Also, though a partial covering model may fit the data well, such a model may be too simplistic as higher signal-to-noise spectra would require more components; we use partial covering as a simple proxy for the complex absorption likely occuring and use this to analyze trends that result from modeling spectra more complicated than a single absorbed power-law.

As Table \ref{cf} shows, varying $\Gamma$ does not significantly alter the results. Though the observed flux for a fixed luminosity does decrease with increasing $\Gamma$, this effect is likely folded into the normalization of the simulated luminosity by the model luminosity at $N_{H,sim}$=0. As expected, $N_{H,sim}$ increases with decreasing covering fraction (Table \ref{gamma}) since the predicted count rate increases with decreasing covering fraction. For the sources predicted to be more heavily obscured ($N_{H,sim} > 10^{24}$ cm$^{-2}$), the predicted column density is not a robust estimate since the absorption-only model we applied does not take into account the effects of Compton scattering. 

To account for Compton scattering, we ran the simulation using the \textit{plcabs} model in XSpec \citep{plcabs}, with the number of scatters ($n_{scatt}$) set equal to 1 for 0 $<$ N$_{H,sim} < 10^{24}$ cm$^{-2}$, $n_{scatt}$=5 for $10^{24}$ cm$^{-2} \leq$ $N_H < 5 \times 10^{24}$ cm$^{-2}$, and $n_{scatt}$=12 for $N_{H,sim} \geq 5 \times 10^{24}$. The results for this simulation are listed in Table \ref{plcabs}. Compared to the most conservative simulation above (i.e. covering fraction of 0.99 and $\Gamma$=1.8), the N$_{H,sim}$ values from the \textit{plcabs} model do not change significantly for the least obscured sources and are reduced by a factor of about 2-3 for the more obscured sources. We fit our sources with the \textit{plcabs} model to compare the fitted N$_H$ values from this model with the simulated values. Excluding 1346+6423 which had an unconstrained \textit{plcabs} fit due to low S/N, we find that 6 of our 16 sources have a fitted N$_H$ an order of magnitude or lower than the simulated values, consistent with the results from comparing the partial covering model with the fitted values from the single or double absorbed power law fits.\footnote{We note, however, that there is a discrepancy between the targets that have under-predicted N$_H$ values using the different models: with the partial covering model, 1437+3634 is underpredicted by an order of magnitude, while the fitted and simulated N$_H$ values are more consistent using the \textit{plcabs} model, and the \textit{plcabs} model under-predicts the column density for 0053-0846, whereas these two values are consistent using the partial covering model. The remaining 5 targets (0325-0608, 0959+1259, 1018+3722, 1111+0228, and 1123+4703) have fitted N$_H$ values underpredicted by both models, which could be due to the non-negligible dispersion in L$_{2-10keV}$/L$_{[OIII]}$: a high simulated L$_{2-10keV}$ value can result in N$_{H,obs}$ $>$ N$_{H,sim}$.} Though the simulations using this model do not require Compton-thick absorption, our main result from the previous simulations still holds: the true column density, as assumed by X-ray attenuation, is under-predicted by an order of magnitude or more for several sources. Also, the \textit{plcabs} model assumes a spherical geometry, rather than the putative torus thought to obscure the central engine. When using a partial covering model, which moves to a more toroidal geometry as the covering fraction decreases, the simulated column density increases and in some cases, becomes Compton-thick.

\begin{figure}[f]
\epsscale{0.60}
\plotone{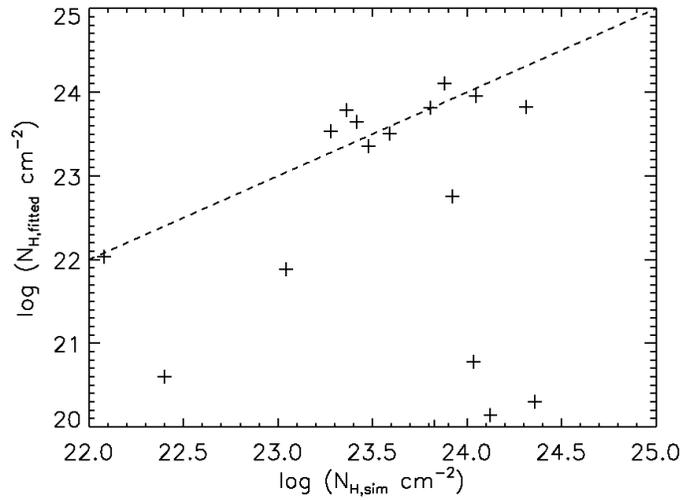}
\caption{\label{nh_fit_nh_sim}Fitted N$_H$ vs. the simulated N$_H$ from the partial covering model with $\Gamma$=1.8 and covering fraction set to 99\%. The dashed line indicates where the two quantities are equal. The simulated column densities are often significantly larger than the fitted values, and are more consistent with the large amounts of inferred X-ray absorption in these sources, as suggested by Figures \ref{hist}, \ref{hist_oiv} and \ref{ews}.}
\end{figure}

In Figure \ref{nh_fit_nh_sim} we plot the simulated values of the column density from the partial covering model with $\Gamma$=1.8 and covering fraction of 99\%  versus those derived from the simple spectral fits. While the values agree reasonably well for many cases, the simulated column densities exceed the fitted values by more than one order-of-magnitude in 6 cases. We note that for 5 of the sources (1136+5657, 1157+5249, 1218+4706, 1238+0927 and 1323+4318), the fitted column density exceeds that of the simulated. However, this disrepancy is by a factor of about 3 or lower in all cases whereas the fitted column density can underpredict the simulated by as much as 4 orders of magnitude (e.g. 0053-0846). The large discrepancy between the fitted and simulated column density reaffirms the importance of using the L$_{2-10 keV}$/L$_{[OIII]}$ ratio (or its equivalent) as a diagnostic of Compton-thickness rather than the fitted column density alone (as also found by Mel\'{e}ndez et al. 2008).

\section{Do Compton-Thick AGN Trace a Distinct Population?}

\begin{figure}[f]
\epsscale{1.08}
\plotone{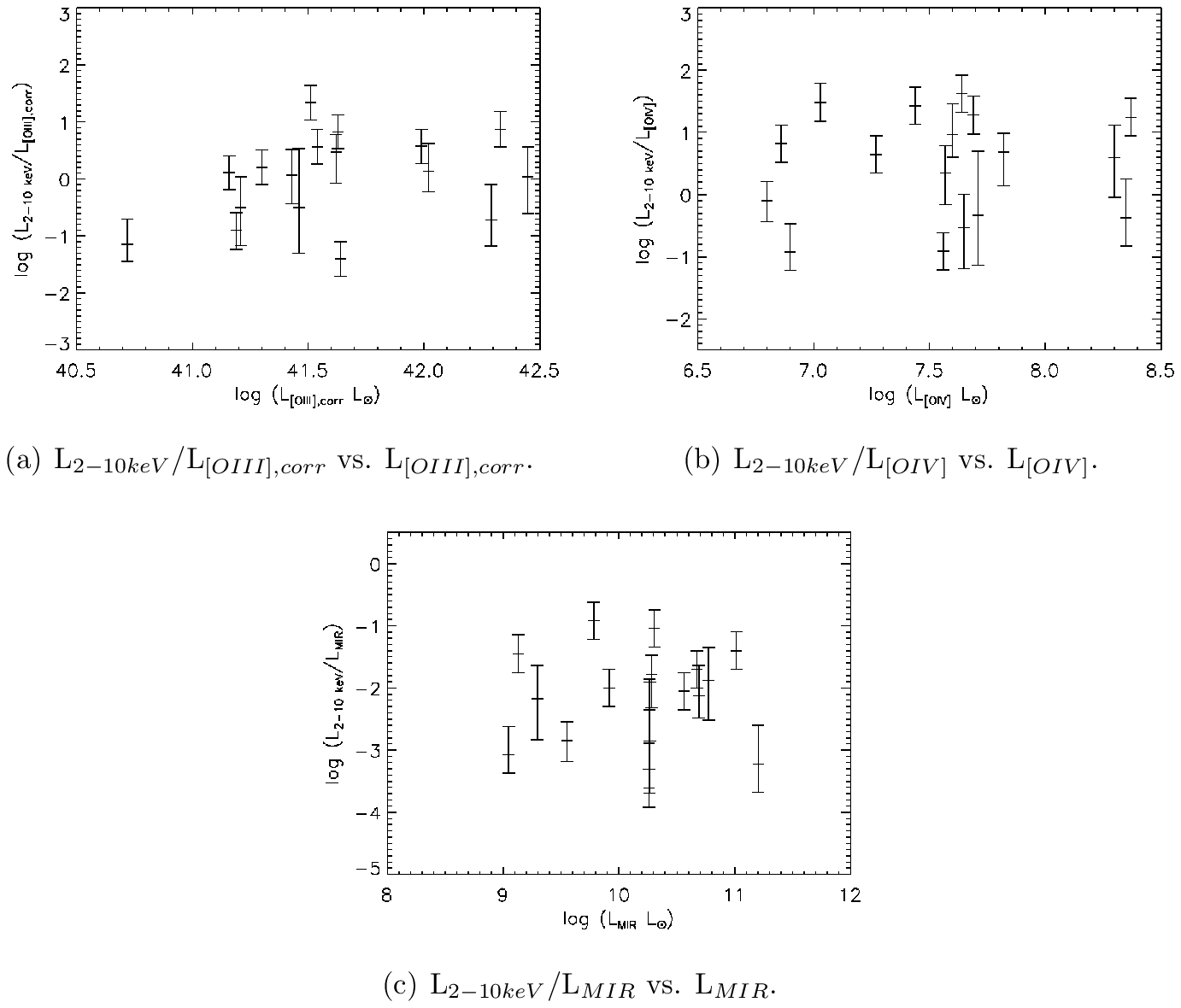}
\caption{\label{lx_agn}Compton-thick diagnostics plotted against proxies of intrinsic AGN luminosity. No correlations are present between the amount of implied X-ray absorption and the AGN intrinsic luminosity.}
\end{figure}

\begin{figure}[f]
\epsscale{1.2}
\plottwo{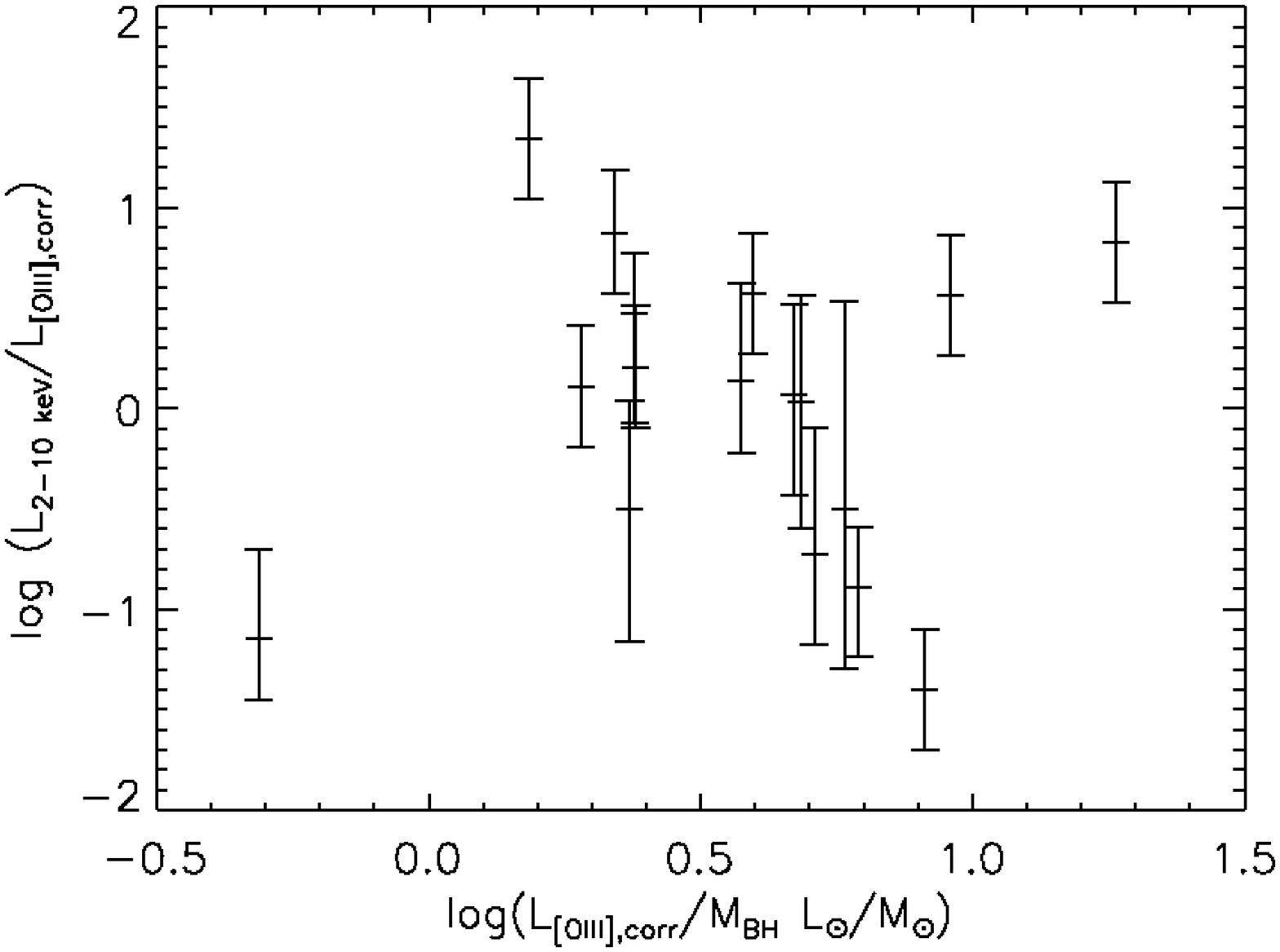}{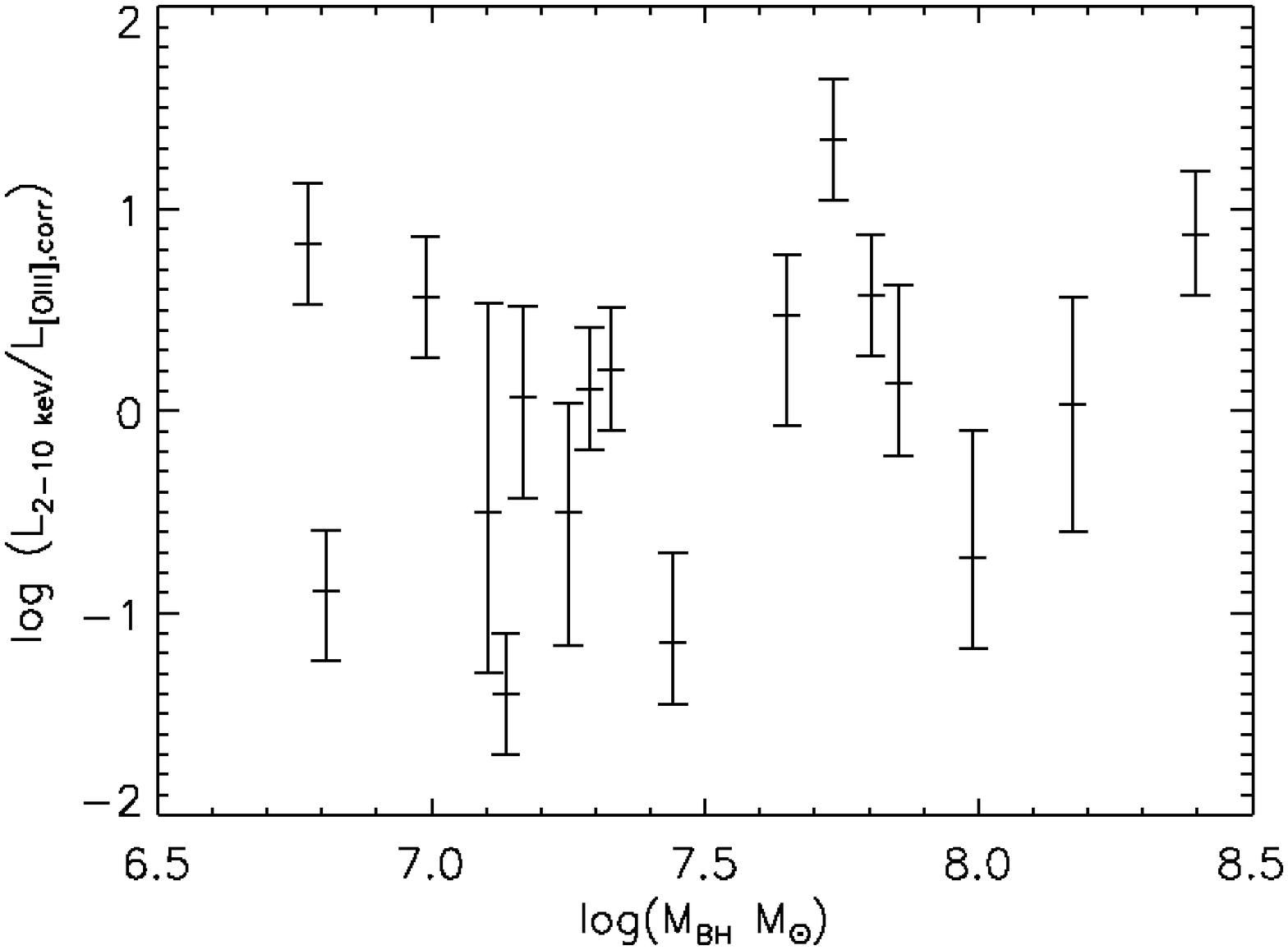}
\caption{\label{ratio_mbh_oiii}Left: L$_{2-10keV}$/L$_{[OIII],corr}$ vs. L$_{[OIII],corr}$/M$_{BH}$, a proxy for the Eddington ratio. There is no correlation between the amount of implied X-ray absorption and the AGN luminosity relative to the Eddington limit. Right: L$_{2-10keV}$/L$_{[OIII],corr}$ vs. M$_{BH}$. No correlation is seen.}
\end{figure}

In this section, we ask whether the Compton-thick AGN population differs systematically from the less obscured objects in other basic properties.
In Figure~\ref{lx_agn} we test whether there is any correlation between the amount of hard X-ray attenuation and the intrinsic luminosity of the AGN (using L$_{[OIII],corr]}$, L$_{[OIV]}$, and L$_{MIR}$). No correlation is seen, at least across the relatively small range in AGN luminosity that we probe in this sample. Similarly, in Figure \ref{ratio_mbh_oiii} (left) we plot the ratio of L$_{2-10keV}/$L$_{[OIII],corr}$ {\it vs.} the AGN luminosity normalized by the black hole mass. As discussed in Kauffmann \& Heckman (2009), this ratio is a rough proxy for the Eddington ratio (with L$/$L$_{Edd} =$ 1 corresponding to L$_{[OIII],corr}/$M$_{BH} \sim$ 50).
Our sample only spans a relatively narrow range in the Eddington ratio, but within this range we see no trend with the amount of X-ray attenuation. We note that previous studies have shown that the X-ray bolometric correction (i.e. ratio of 2-10 keV X-ray luminosity to bolometric luminosity) increases with the Eddington ratio (e.g. Vasudevan \& Fabian 2009).  Likewise, no correlation is seen if we replace L$_{[OIII],corr}$ with L$_{OIV}$ or L$_{MIR}$. Finally, in Figure \ref{ratio_mbh_oiii} (right) we plot L$_{2-10keV}/$L$_{[OIII],corr}$ {\it vs.} the black hole mass. Again, no correlation is present. 

\begin{figure}[f]
\epsscale{1.2}
\plottwo{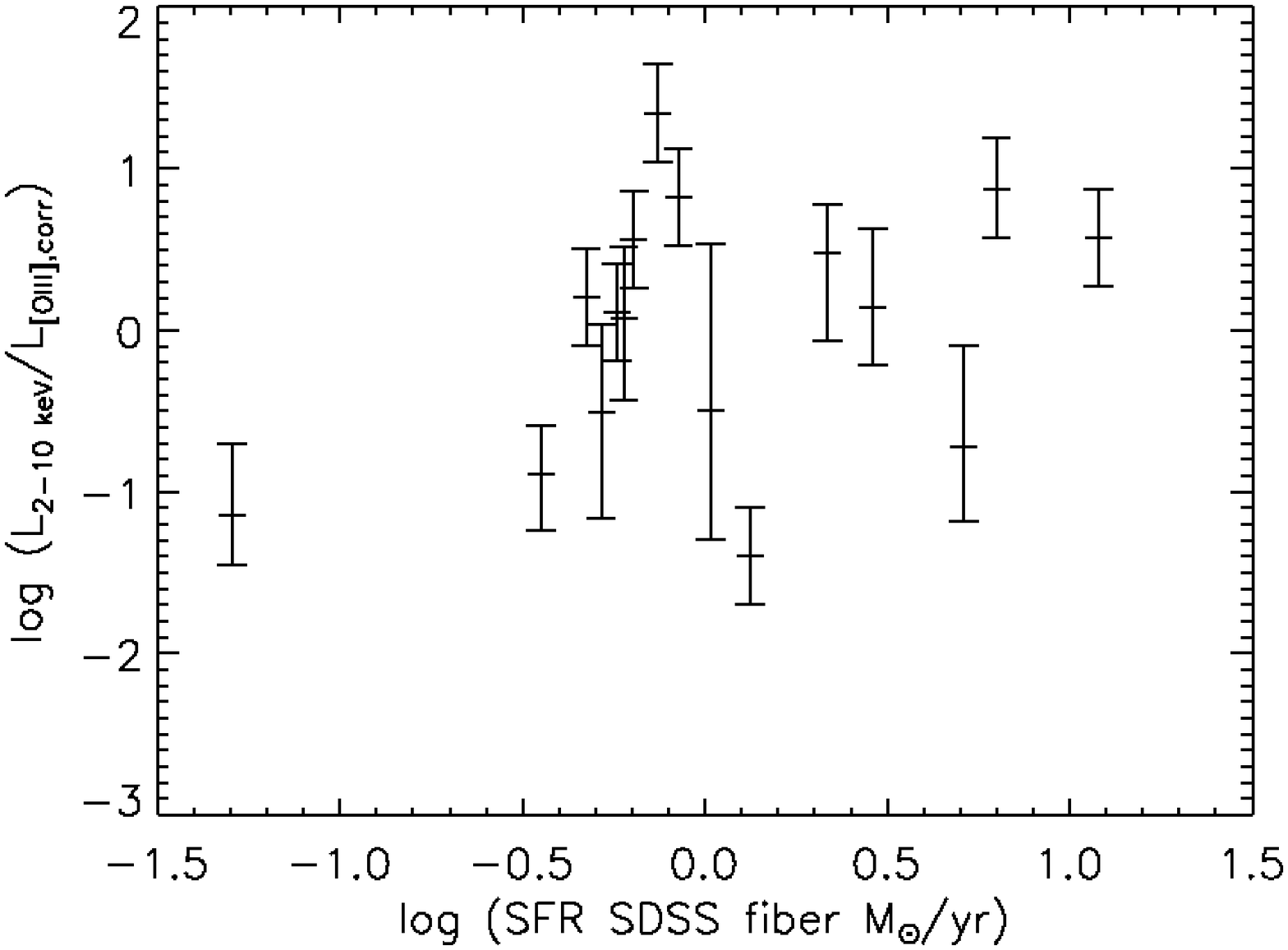}{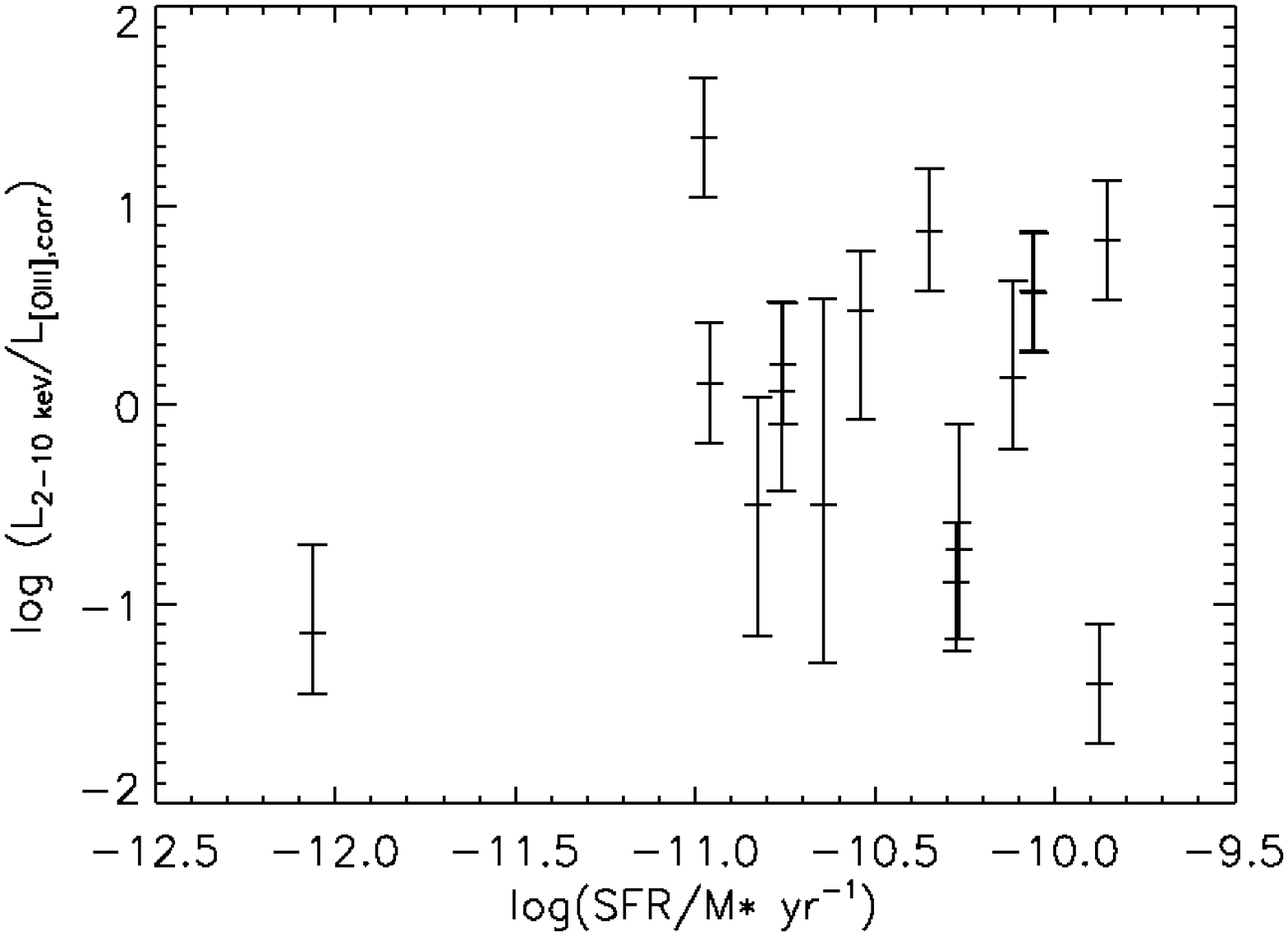}
\caption{\label{lx_sfr}Left: L$_{2-10keV}$/L$_{[OIII],corr}$ vs. SFR in the SDSS fiber. Right: L$_{2-10keV}$/L$_{[OIII]}$ vs. SFR in the SDSS fiber divided by M*. In both plots, no correlation is seen, suggesting that the gas responsible for the X-ray absorption is not directly related to the gas that fuels the star formation.}
\end{figure}

Some of the best-known Compton-thick AGN are found in nuclei that are also undergoing intense bursts of star-formation (e.g. Levenson et al. 2004, 2005). It would be plausible that there is some physical connection between the high gas column densities necessary for such starbursts and high column densities in the X-ray obscuring material. To investigate this possibility in our sample, we show in the left panel of Figure \ref{lx_sfr} a plot of L$_{2-10keV}/$L$_{[OIII],corr}$ {\it vs.} the star-formation rate within the SDSS spectroscopic fiber (typical projected size of a few kpc in our sample). In the right panel of Figure \ref{lx_sfr} we plot L$_{2-10keV}/$L$_{[OIII],corr}$ {\it vs.} the star-formation rate per unit stellar mass (SFR/$M_*$). Neither of these figures shows any correlation between the amount of X-ray obscuration and the amount (relative or absolute) of circum-nuclear star formation. The lack of correlation between obscuration and host galaxy properties indicates that the obscuration affecting X-ray emission is likely due to the torus rather than the large scale ISM of the host galaxy. As our sample was selected on the basis of bright [OIII] emission, this could be in part a selection effect against AGN with large amounts of extinction produced by the host galaxy (which would affect the [OIII] flux).

\section{Obscuration at Higher Energies}

We have shown that the amount of X-ray obscuration in the 2-10 keV band is highly significant for optically selected Type 2 AGN. Over half of the sources have attenuations exceeding an order-of-magnitude, implying absorbing column densities in excess of $10^{24}$ cm$^{-2}$. Do X-ray surveys at higher energies recover these highly obscured AGN? To investigate this, we compare our results to those for samples of AGN detected in the \textit{Swift} BAT survey in the 15-195 keV band: the Landi et al. (2007) sample, based on AGN identification of previously unidentified \textit{Swift} sources (4 Sy2s); the Winter et al. (2008) sample, based on \textit{XMM-Newton} observations of AGN detected by \textit{Swift} with high significance (10 Sy2s with $z$ = 0.01 - 0.09, though only 4 had published $f_{[OIII]}$, $f_{H\alpha}$ and $f_{H\beta}$ values, see below); and the radio quiet Type 1.8, 1.9, and 2 Seyfert galaxies in the Mel\'{e}ndez et al. (2008) sample, based on the first 3 months of the \textit{Swift} BAT high Galactic latitude survey (14 targets with $z$ = 0.001 - 0.03, complete to (1-3) $\times 10^{-11}$ erg cm$^{-2}$ s$^{-1}$). 

First, we use the extinction-corrected luminosity of the [OIII] line as our proxy for the intrinsic luminosity. To do so, we use the published L$_{2-10 keV}$ values from Winter et al. (2008) and the published L$_{2-10 keV}$ and [OIII] flux values ($f_{[OIII]}$) from the Landi et al. (2007) and Mel\'{e}ndez et al. 2008 samples (restricted to just the radio-quiet Type 1.8, 1.9 and 2 Seyfert galaxies). We also use the Landi et al.  H$\alpha$ ($f_{H\alpha}$) and H$\beta$ ($f_{H\beta}$) flux values to correct for extinction. For the Winter et al. and Mel\'{e}ndez et al. 2008 samples, we used $f_{[OIII]}$, $f_{H\alpha}$ and $f_{H\beta}$ values from the literature where available and include only those sources that have published flux values for all 3 parameters.\footnote{Swift J1200.8+0650: $f_{[OIII]}$, $f_{H\alpha}$, and $f_{H\beta}$ values from Landi et al. (2007). NGC 1142 and Mrk 417: $f_{[OIII]}$, $f_{H\alpha}$, and $f_{H\beta}$ values from Dahari \& De Robertis 1988. Swift J1200.8+0650 was in both the Landi et al. and Winter et al. paper. Here we use the L$_{2-10 keV}$ from the Winter et al. paper, but utilize the Landi et al. paper for the optical spectral fluxes.For Mel\'{e}ndez et al. sample (2008), $f_{[OIII]}$ values from Mel\'{e}ndez et al. (2008) and $f_{H\alpha}$ and $f_{H\beta}$ values from: Bassani et al. 1999, Dahari \& De Robertis 1988, Polletta et al. 1996, and Tran 2003.} In some cases, the published $f_{H\beta}$ values were upper limits, providing a lower limit on L$_{[OIII],corr}$ and an upper limit on the L$_{2-10 keV}$/L$_{[OIII],corr}$ ratio.

\begin{figure}[f]
\epsscale{1.2}
\plottwo{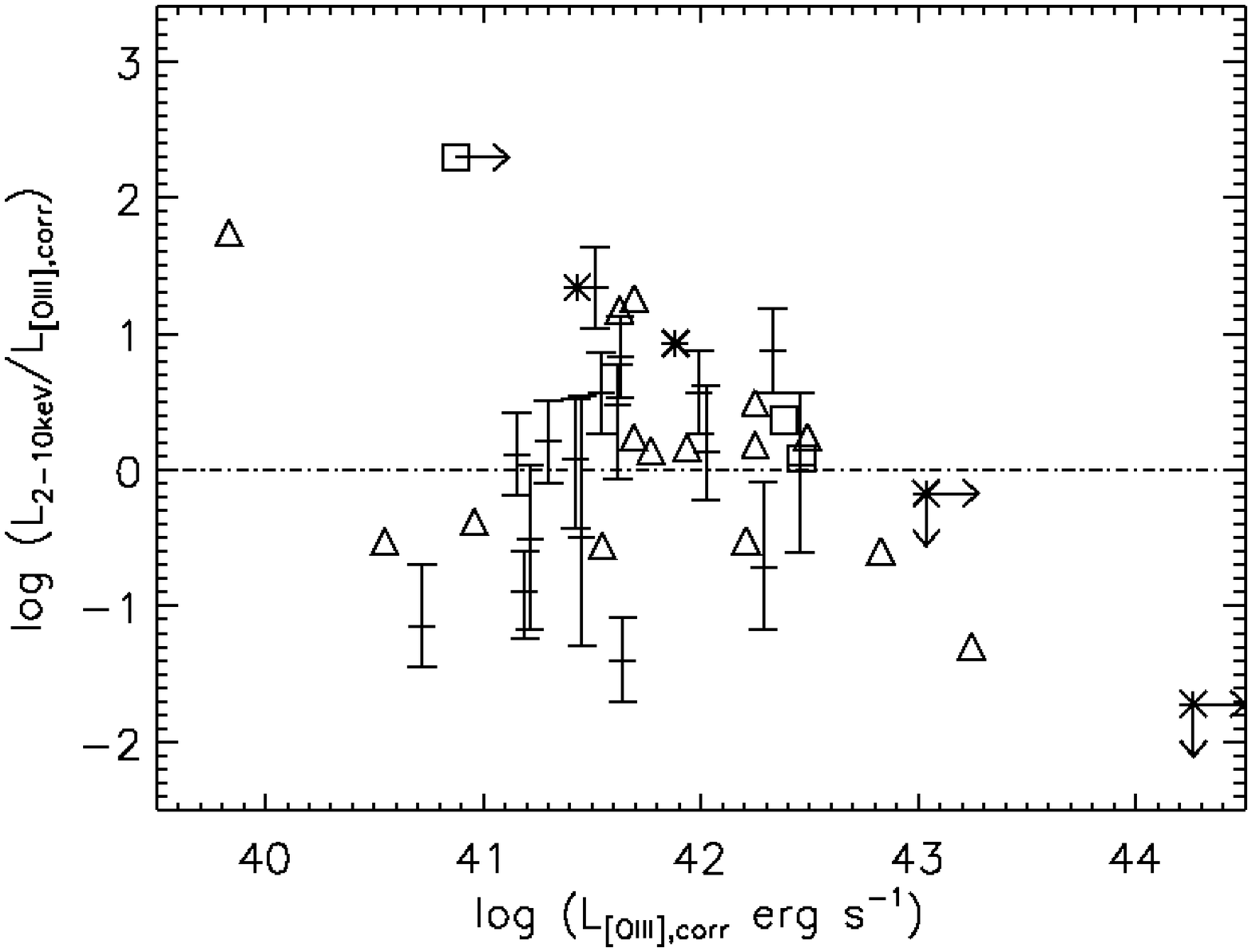}{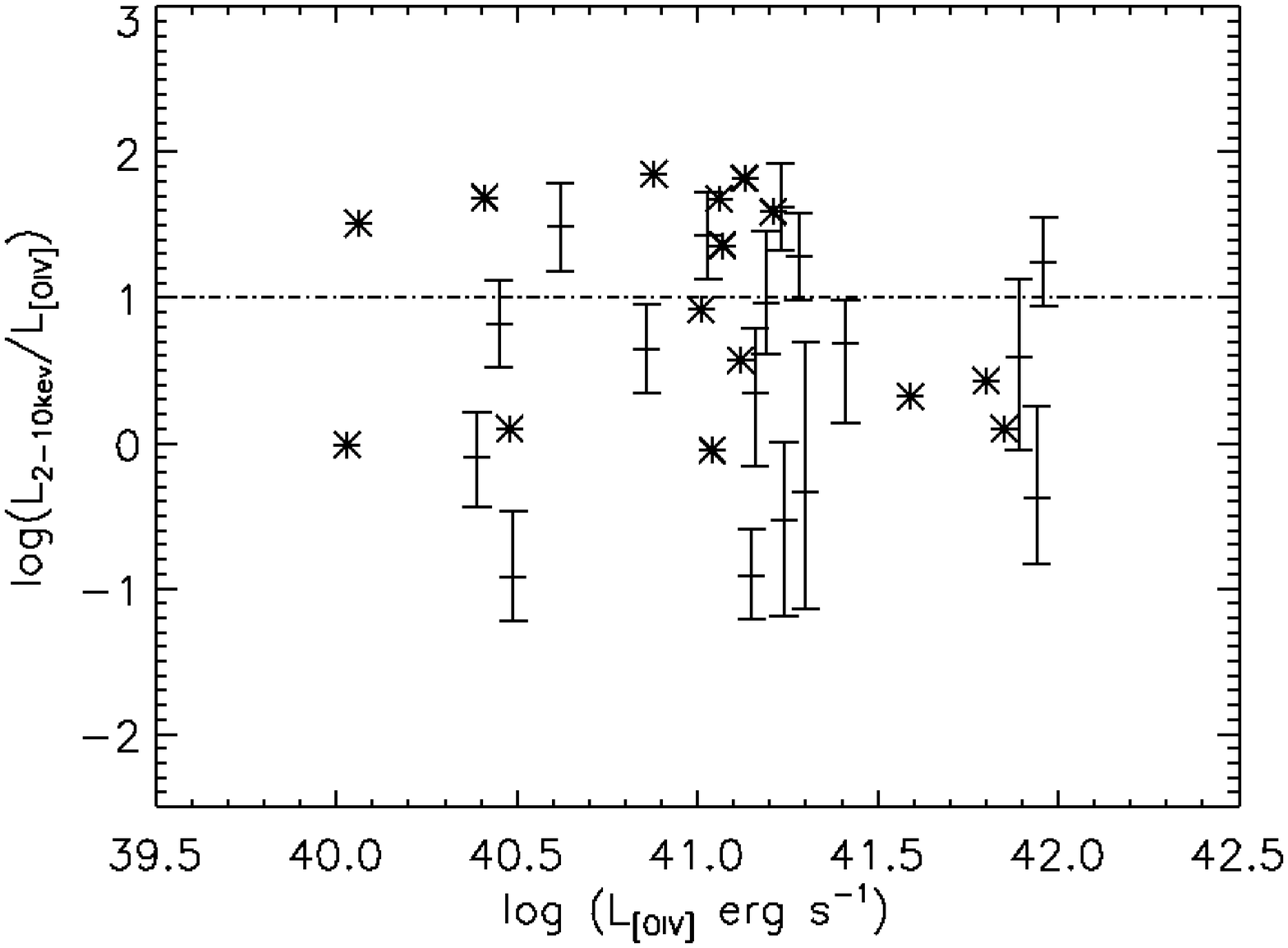}
\caption{\label{other_samp} Left: L$_{2-10 keV}$/L$_{[OIII],corr}$ vs. extinction-corrected L$_{[OIII]}$ for our sample and three other samples selected from the \textit{Swift} survey, in the 15-195 keV band. The Landi et al. (2007) targets are marked by asterisks, the Winter et al. sample is shown by the squares (2008) and the triangles indicate the Sy 1.8, 1.9 and 2 galaxies from Mel\'{e}ndez et al. sample (2008). Right: L$_{2-10keV}$/L$_{[OIV]}$ vs L$_{[OIV]}$ for our sample and the Sy 1.8, 1.9 and 2 galaxies from Mel\'{e}ndez et al. sample (2008), marked by asterisks. In both plots, the dashed-dotted line indicates the ``nominal'' Compton-thick/Compton-thin boundary. These samples of AGN show distributions in the ratios of L$_{2-10 keV}$/L$_{[OIII],corr}$ and L$_{2-10keV}$/L$_{[OIV]}$ that are similar to those of our sample. This suggests that selection in the 15-195 keV band is recovering a significant fraction of the Compton-thick population seen in our sample.}  
\end{figure}

A plot of L$_{2-10 keV}$/L$_{[OIII],corr}$ vs L$_{[OIII],corr}$ for our sample, the Landi et al. sample (asterisks), the Winter et al. sample (squares) and the Mel\'{e}ndez et al. sample (triangles) is shown in Figure \ref{other_samp} (left). The dashed-dotted line indicates the nominal boundary between Compton-thick and Compton-thin candidates (i.e. L$_{2-10 keV}$/L$_{[OIII],corr}$ = 1). It is clear that these BAT-selected source show a very large range in L$_{2-10 keV}$/L$_{[OIII],corr}$, similar to that seen in our optically selected sample. This can be contrasted with the distribution of L$_{2-10 keV}$/L$_{[OIII]}$ for AGN selected in the softer 3-20 keV band (from the Sazonov \& Revnivstev 2004 sample), as found by Heckman et al. (2005): the distribution is relatively narrow and there is no significant difference between the Type 1 and Type 2 AGN. This is because such samples exclude the Compton-thick Type 2 AGN.

Next, we use the [OIV] 25.89$\mu$m line luminosity as a tracer of the AGN luminosity. In the right panel of Figure \ref{other_samp} we compare the distribution of log (L$_{2-10 keV}/$L$_{[OIV]}$ as function of log (L$_{[OIV]}$) for our optically-selected sample to that of the BAT-selected sample of radio-quiet Type 2 AGN in Mel\'{e}ndez et al. (2008). The distributions are similar, with 8/15 of the BAT-selected objects and 10/17 of our optically-selected objects having ratios of L$_{2-10 keV}/$L$_{[OIV]}$ more than an order-of-magnitude lower than the mean value for radio-quiet Type 1 AGN in Mel\'{e}ndez et al (log (L$_{2-10 keV}/$L$_{[OIV]}$) $\sim$ 2.1 $\pm$ 0.3 dex). 

Taken together, the results in Figure \ref{other_samp} imply that selection in the 14-195 keV band is recovering a significant fraction of the sources that are heavily obscured in the 2-10 keV band. This agrees at least qualitatively with the conclusions of Winter et al. (2009): of 22 AGN in the sample detected in the 14-195 keV band, 5 are Compton-thick candidates and an additional 4 were hidden/buried, or very Compton-thick, AGN that would be missed in sample selection at lower energies. However, this contrasts with the results of Treister et al. (2009) who found a  low number of Compton-thick AGN using high energy (E $>$ 10 keV) surveys: 5 of the 130 AGN detected with INTEGRAL and 8 of the 130 AGN detected with Swift were Compton-thick.

\section{Conclusions}

In this paper we have undertaken a study of the hard X-ray properties (2-10 keV) of a nearly complete sample of the 17 brightest Type 2 (obscured) AGN selected from a sample of 480,000 galaxies in the Fourth Data Release of the Sloan Digital Sky Survey. These are representative of optically-bright Type 2 AGN in the low-redshift universe ($z <$ 0.l). 

We have used the value of the luminosity of the [OIII]$\lambda$5007 emission-line for the AGN's Narrow Line Region (NLR) -corrected for dust extinction using the observed H$\alpha$/H$\beta$ - as a proxy for the intrinsic luminosity of the AGN (e.g. Bassani et al. 1999; Kauffmann \& Heckman 2009). We have also used the {\it Spitzer Space Telescope} to measure the luminosity of the mid-infrared [OIV] 25.89$\mu$m line from the NLR (e.g. Mel\'{e}ndez et al. 2008) and the luminosity of the mid-infrared continuum from the obscuring torus (e.g. Nenkova et al. 2008) as additional indicators of the AGN intrinsic luminosity. The [OIII] and [OIV] lines are useful because they originate well outside the torus and should therefore be unaffected by its obscuration. In contrast, the hard X-rays originate inside the torus and can be strongly attenuated in Type 2 AGN for a Compton-thick torus with N$_H > 10^{24}$ cm$^{-2}$. We have therefore used the ratio of the observed hard X-ray luminosity to the luminosities of the [OIII] and [OIV] emission-lines and to the mid-infrared continuum luminosity as an indicator of the amount of obscuration of the hard X-ray source. We have also used the equivalent width of the Fe K$\alpha$ emission line at 6.4 keV as an additional diagnostic of a heavily obscured hard X-ray source.

We found that the ratio of the hard X-ray luminosity to that of the [OIII] and [OIV] lines and of the mid-infrared continuum in our sample spans an enormous range (nearly four orders-of-magnitude). In a majority (11/17) of the cases, the luminosity ratios are more than an order-of-magnitude lower than the mean value for Type 1 (unobscured) AGN, and these objects are likely to be Compton-thick. While similar results have been reported before (e.g. Bassani et al. 1999; Heckman et al. 2005), our results are the first that have been determined for a complete sample of Type 2 AGN selected in homogeneous way on the basis of their optical emission-line flux.

In principle, the relative weakness of the hard X-ray emission could be an intrinsic property of the AGN, rather than a consequence of obscuration (e.g. Brightman \& Nandra 2008). We have used the properties of the Fe K$\alpha$ line at 6.4 keV to distinguish between these possibilities. We found that there is a good inverse correlation in our sample between the equivalent width of the K$\alpha$ line and relative luminosity of the 2-10 keV continuum. We have argued that this is consistent with an obscured X-ray source in which the emergent spectrum has been significantly reprocessed. In such a case the K$\alpha$ photons traverse significantly lower column densities of absorbing gas than do the underlying ($\sim$ 6 to 7 keV) continuum photons, and large K$\alpha$ equivalent widths result (e.g. Krolik \& Kallman 1987; Levenson et al. 2006). Our results are consistent with those reported by Bassani et al. (1999) which were based on a heterogeneous sample of Type 2 AGN.

We have shown that in some cases the high amount of inferred obscuration is inconsistent with the estimate of the column of absorbing gas derived from fits of simple spectral models to the data (e.g. single or double power-law spectra transmitted through a homogeneous foreground screen). We have constructed more physically realistic models with partial covering of the X-ray source plus Compton scattering. We run simulations of these models constrained by the ratio of the 2-10 keV and [OIII]$\lambda$5007 fluxes and found that in some cases significantly higher absorbing column densities were implied than those derived from the simple spectral fits. We have concluded that the latter columns are not reliable for heavily obscured (Compton-thick) AGN. Similar results have been obtained by Mel\'{e}ndez et al. (2008).

We found that the populations of Compton-thick and Compton-thin Type 2 AGN in our sample do not differ systematically in terms of the luminosity of the AGN, the mass of the black hole, or the luminosity with respect to the Eddington limit. Likewise, we found no differences in the star formation rate or the star formation rate per unit stellar mass (measured in the central few-kpc-scale region covered by the SDSS spectra).

We have also compared our results for an optically-selected sample of Type 2 AGN to those for samples selected in the {\it Swift} BAT 14-195 keV energy band (Landi et al. 2007; Winter et al. 2008; Mel\'{e}ndez et al. 2008). Using the relative strength of the 2-10 keV and the [OIII]$\lambda$5007 and [OIV] 25.89 $\mu$m luminosities, we find that the BAT-selected AGN have an implied distribution of attenuation in the 2-10 keV band that is similar to that seen in our optically-selected sample. This suggests that samples selected in the {\it Swift} BAT band do recover a significant fraction of the local population of Compton-thick AGN.

We conclude that, despite the penetrating power of hard X-rays, the majority of low-redshift Type 2 AGN still suffer significant obscuration in the 2-10 keV band. Thus, existing X-ray surveys in this band will have missed a significant fraction of the population of obscured AGN. Fortunately, it appears that the missing population of X-ray-selected AGN do not differ systematically in terms of the basic properties of the AGN and of the stellar populations in their host galaxy. That is - at least in the local universe - the Compton-thick and Compton-thin Type 2 AGN do not seem to trace distinct populations.

\acknowledgments{This work was funded by NASA grant number NNX07AQ36G. The authors thank the anonymous referee for insightful comments and suggestions which improved the quality of the manuscript.}

\clearpage

\clearpage

\begin{table}[h]
\small
\caption{\label{sample}Sample and \textit{XMM-Newton} Observation Log}
\begin{tabular}{lclccc}
\hline
\hline\noalign{\smallskip}
Target & Distance$^1$ & Alternate Name  & ObsID & Observation Start Date & Exposure Time$^2$ \\
& &  & & & MOS1/MOS2/PN \\
&       Mpc & & & UT & ks \\
\hline\noalign{\smallskip}

0053-0846 & 81.4  & NGC 0291  & 0504100301 & 2007-06-24 & 6.4/6.2/2.2 \\

0325-0608 & 147.4 & Mrk 609   & 0103861001 & 2002-08-13 & 9.1/9.1/7.5 \\

0800+2636 & 116.4 & IC 0486   & 0504101201 & 2007-10-28 & 21.6/21.6/20.1\\

0804+2345 & 125.3 & 2MASX J0804+2345 & 0504102101 & 2007-11-09 & 22.6/22.6/20.4\\

0824+2959 & 107.6 & IRAS F08216+3009  & 0504102001 & 	2007-11-03 & 23.4/23.4/21.8 \\

0959+1259 & 147.4 & CGCG 064-017 & 0504100201 & 2007-11-25 & 21.6/21.6/20.0 \\

1018+3722 & 214.9 & 2MASX J1018+3722  & 0504101701 & 2007-05-18 & 23.2/23.2/18.8 \\

1111+0228 & 151.9 & 2MASX J1111+0228 & 0504101801 & 2007-06-10 &  24.0/24.0/22.6 \\

1123+4703 & 107.6 & CGCG 242-028  & 0504101301 & 2007-06-11 & 19.5/19.8/4.0 \\

1136+5657 & 224.0 & MCG +10-17-021  & 0504101001 & 2007-12-15 & 21.6/21.6/19.8 \\

1147+5226 & 214.9 & Mrk 1457 & 0504101401 & 2008-04-21 & 26.4/26.4/14.5 \\

1157+5249 & 156.3 & 2MASX J1157+5249 & 0504100901 & 2007-06-21 & 25.0/25.2/22.1 \\

1218+4706 & 425.8 & 2MASX J1218+4706 & 0203270201 & 2004-06-01 & 43.7/43.9/36.6 \\

1238+0927 & 373.0 & 2MASX J1238+0927 & 0504100601 & 2007-12-09 & 21.6/21.6/20.1 \\

1323+4318 & 116.4 & CGCG 218-007 & 0504101601 & 2007-11-21 & 8.3/8.3/5.7 \\

1346+6423 & 103.3 & 2MASX J1346+6423 & 0504101501 & 2007-06-13 & 10.5/14.0/1.3 \\

1437+3634 & 59.8  & NGC 5695 & 0504100401 & 2007-12-16 & 20.4/20.2/11.2 \\

\hline
\hline
\multicolumn{6}{l}{$^1$Distances based on optical spectroscopic redshift using H$_o$=71 km s$^{-1}$, $\Omega_{M}$ = 0.27 and $\Omega_{\Lambda}$ = 0.73.}\\
\multicolumn{6}{l}{$^2$Net exposure time after filtering.}
 \end{tabular}
\end{table}

\begin{table}[h]
\small
\caption{\label{counts}X-ray Counts$^1$}
\begin{tabular}{lrr}
\hline
\hline\noalign{\smallskip}
Target & Soft & Hard \\
       & 0.5-2 keV & 2-10 keV\\      
\hline\noalign{\smallskip}

0053-0846 & 62   & 18   \\ 

0325-0608 & 5392 & 1898 \\

0800+2636 & 5327 & 11477 \\ 

0804+2345 & 163  & 67   \\

0824+2959 & 816  & 2188 \\

0959+1259 & 2858 & 3327 \\

1018+3722$^2$ &  233  & 19/$<$34/$<$38 \\

1111+0228 & 304  & 416  \\

1123+4703 & 63   & 59   \\

1136+5657 & 474  & 81   \\

1147+5226 & 261  & 316  \\

1157+5249 & 530  & 179  \\

1218+4706 & 76   & 83   \\

1238+0927 & 658  & 1934 \\

1323+4318 &  100  & 506  \\

1346+6423 & 16   & 24   \\

1437+3634$^2$ & 299  & 71/$<$40/15 \\

\hline
\hline
\multicolumn{3}{l}{$^1$About half the total counts are from the PN}\\
\multicolumn{3}{l}{detector with the remainder roughly equally}\\
\multicolumn{3}{l}{split between the MOS1 and MOS2 detectors.}\\
\multicolumn{3}{l}{$^2$In the 2-10 keV band, there are 3$\sigma$ upper limits}\\
\multicolumn{3}{l}{count rate for one or both of the MOS detectors.}\\
\multicolumn{3}{l}{Counts are therefore listed separately for each}\\
\multicolumn{3}{l}{detector: PN/MOS1/MOS2}\\
\end{tabular}
\end{table}

\begin{table}[h]
\small
\caption{\label{pow}Parameters for Powerlaw Fits}
\begin{tabular}{lcccccr}
\hline
\hline\noalign{\smallskip}
Target & N$_H$ & $\Gamma_1$ & N$_H$ & $\Gamma_2$ & $\chi^2$  & $\chi^2$ 1pow\\
&        10$^{22}$ cm$^{-2}$ &    &  10$^{22}$  cm$^{-2}$ &  & (DOF) & (DOF)\\
\hline\noalign{\smallskip}

0325-0608*   & 0.04 & 1.84$^{+0.03}_{-0.03}$ & & &  & 173.6 (205) \\

0800+2636*   & 0.03 & 1.25$^{+0.07}_{-0.06}$ & 1.08$^{+0.09}_{-0.08}$ & =$\Gamma_1$ & 495.9 (448) & 758.1 (449) \\

0804+2345    & $0.28^{+0.17}_{-0.23}$ & $4.66^{+2.63}_{-1.70}$ & $67.0^{+30.0}_{-30.4}$ &  =$\Gamma_1$ &
 28.4 (42) & 41.4 (44)\\

0824+2959*  & 0.03 & 2.72$^{+0.12}_{-0.11}$ & 22.9$^{+2.3}_{-1.7}$ & =$\Gamma_1$ & 227.1 (165) & 924 (169)\\

0959+1259    & $0.77^{+0.05}_{-0.05}$ & $1.88^{+0.08}_{-0.08}$ & & & 
 & 205.7 (267)\\

1018+3722${^1}$  & 0.01 & 3.17$^{+0.38}_{-0.33}$ & & & & 72.9 (63) \\

1111+0228*   & 0.04 & 1.97$^{+0.25}_{-0.25}$ & 5.75$^{+1.76}_{-1.27}$ & =$\Gamma_1$  & 58.6 (40) & 111.9 (42)\\

1136+5657    & $0.07^{+0.07}_{-0.05}$ & $3.30^{+0.54}_{-0.45}$ & $61.2^{+66.8}_{-28.3}$ & =$\Gamma_1$ & 
 40.8 (50) & 57.4 (52)\\

1147+5226    & $0.11^{+0.09}_{-0.04}$ & $3.36^{+1.09}_{-0.82}$ & $32.1^{+6.5}_{-4.9}$ & =$\Gamma_1$ &
  52.7 (37) & 191.7 (39)\\

1157+5249    & $0.06^{+0.05}_{-0.05}$ & $3.10^{+0.51}_{-0.41}$ & $127^{+38}_{-42}$ & =$\Gamma_1$ & 
 98.4 (57) & 159.1 (59)\\

1218+4706*   & 0.02 & 2.87$^{+0.38}_{-0.43}$ & 89.6$^{+51.4}_{-38.3}$ & =$\Gamma_1$ &  21.1 (21) & 54.5 (23) \\

1238+0927    & 0.03$^{+0.05}_{-0.01}$ & 2.60$^{+0.38}_{-0.23}$ & 33.8$^{+3.8}_{-3.2}$ & =$\Gamma_1$ & 
 179.6 (124) & 777.3 (126) \\

1323+4318    & 0.06$^{+0.34}_{-0.04}$ & 2.74$^{+2.40}_{-0.67}$ & 44.5$^{+12.5}_{-10.3}$ & =$\Gamma_1$ & 
57.6 (64) & 140.6 (68)\\

1437+3634    & $0.06^{+0.24}_{-0.05}$ & 3.01$^{+2.16}_{-0.70}$  & & & & 79.7(64) \\

\hline\noalign{\smallskip}
Target & N$_H$ & $\Gamma_1$ & N$_H$ & $\Gamma_2$ & c-stat & c-stat 1pow\\
&        10$^{22}$ cm$^{-2}$ &    &  10$^{22}$ cm$^{-2}$   & & (DOF) & (DOF)\\
\hline\noalign{\smallskip}

0053-0846    & $0.09^{+0.05}_{-0.09}$ & $2.98^{+1.35}_{-0.87}$ & $64.9^{+51.1}_{-34.4}$ & =$\Gamma_1$ &
43.8 (41) & 57.5 (43)\\

1123+4703*   & 0.01 & 0.83$^{+0.25}_{-0.24}$ & & & & 113.2 (93) \\

1346+6423$^2$   & 0.02 & 1.8 & & & & 77.6 (49) \\

\hline
\hline
\multicolumn{7}{l}{*Best fit absorption same as Galactic absorption.}\\
\multicolumn{7}{l}{$^1$Absorption frozen at Galactic value to fit $\Gamma$ at a physical value.}\\
\multicolumn{7}{l}{$^2\Gamma$ frozen at 1.8 and absorption frozen at Galactic value to constrain spectral fit.}\\
 \end{tabular}
\end{table}

\begin{table}[h]
\small
\caption{\label{ftest} F-test on powerlaw fits}
\begin{tabular}{lcc}
\hline
\hline\noalign{\smallskip}
Target &  F$_{calc}$ & Probability\\
\hline\noalign{\smallskip}

0053-0846 &  6.412   & $>$0.996 \\

0800+2636 &  236.9  & 1\\

0804+2345 &  9.613  & $>$0.9973\\

0824+2959 &  126.6  & 1 \\

1111+0228 &  18.19  &  $>$0.9973 \\

1136+5657 &  10.17  & $>$0.9973 \\

1147+5226 &  48.80  & 1\\

1157+5249 &  17.58  & $>$0.9973 \\

1218+4706 &  16.62  & $>$0.9973 \\

1238+0927 &  206.3  & 1 \\

1323+4318 &  23.1   & 1 \\

\hline
\hline
 \end{tabular}
\end{table}

\begin{table}[h]
\small
\caption{\label{apec}Parameters for Thermal +  Powerlaw Fits}
\begin{tabular}{lccccccr}
\hline
\hline\noalign{\smallskip}
Target & N$_H$ & kT & abund & $\Gamma_1$ & N$_H$ & $\Gamma_2$  & $\chi^2$ \\
&        10$^{22}$ cm$^{-2}$ & keV &  &  &  10$^{22}$ cm$^{-2}$ & &  (DOF) \\
\hline\noalign{\smallskip}

0325-0608* & 0.04 & 0.27$^{+0.05}_{-0.04}$ & 1 & 1.77$^{+0.05}_{-0.04}$ & & & 159.7 (203) \\

0800+2636 & 0.06$^{+0.09}_{-0.03}$ & $<$0.17 & 1 & 1.22$^{+0.08}_{-0.07}$ & 1.0$^{+0.1}_{-0.1}$ & =$\Gamma_1$ &  489.2 (445) \\

0804+2345 & 0.53$^{+0.15}_{-0.37}$ & $<$0.12 & 1 & 5.16$^{+0.99}_{-3.70}$ & 19.7$^{+164.3}_{-17.3}$ & 1.8 & 23.9 (40) \\

0824+2959* & 0.03 & 0.18$^{+0.04}_{-0.05}$ & 1 & 2.49$^{+0.18}_{-0.19}$ & 22.1$^{+2.3}_{-1.9}$ & =$\Gamma_1$ & 214.9 (163) \\

0959+1259 & 0.80$^{+0.07}_{-0.07}$ & - & 1 & 1.90$^{+0.10}_{-0.09}$ & & & 204.4 (265) \\

1018+3722 & 0.12$^{+0.26}_{-0.11}$ & $<$0.20 & 1 & 3.41$^{+2.35}_{-1.31}$ & & & 51.1 (60) \\

1111+0228 & 0.06$^{+0.09}_{-0.02}$ & 0.22$^{+0.07}_{-0.06}$ & 1 & 1.72$^{+0.86}_{-0.65}$ & 
5.2$^{+1.8}_{-1.3}$ & =$\Gamma_1$ & 54.3(37)  \\

1136+5657 & 0.03$^{+0.08}_{-0.02}$ & 0.83$^{+0.23}_{-0.20}$ & 1 & 3.06$^{+0.38}_{-0.30}$ &
 37.9$^{+49.1}_{-21.3}$ & 1.8 & 36.6 (48) \\

1147+5226 & 0.53$^{+0.14}_{-0.12}$ & 0.14$^{+0.03}_{-0.03}$ & 1 & 1.64$^{+0.68}_{-1.00}$ & 27.6$^{+10.3}_{-7.0}$ & =$\Gamma_1$ &
 35.3(35) \\

1157+5249 & 0.05$^{+0.07}_{-0.03}$ & 0.20$^{+0.05}_{-0.03}$ & 1 & 2.69$^{+0.51}_{-0.38}$ & 123$^{+54}_{-44}$ & =$\Gamma_1$ &
 79.5(55) \\

1218+4706* & 0.02 & $<$0.24 & 1 & 1.96$^{+0.70}_{-0.86}$ & 90.3$^{+62.7}_{-36.8}$ & =$\Gamma_1$ & 15.0 (19) \\

1238+0927 & 0.03$^{+0.06}_{-0.02}$ & $<$0.22 & 1 & 2.37$^{+0.45}_{-0.28}$ & 32.6$^{+4.1}_{-3.1}$ & =$\Gamma_1$ & 167.6 (122) \\

1323+4318 & 0.02$^{+0.18}_{-0.01}$ & - & 1 & 2.55$^{+1.64}_{-0.51}$ &  43.7$^{+4.7}_{-7.6}$ & =$\Gamma_1$ & 56.0 (62) \\

1437+3634* & 0.01 & 0.25$^{+0.74}_{-0.13}$ & 1 & 2.41$^{+0.60}_{-0.39}$ & & & 75.8 (63)\\

\hline\noalign{\smallskip}
Target & N$_H$ & kT & abund &  $\Gamma_1$ & N$_H$ & $\Gamma_2$ &  c-stat \\
&        10$^{22}$ cm$^{-2}$ &  keV & &  &  10$^{22}$ cm$^{-2}$ & &  (DOF) \\
\hline\noalign{\smallskip}

0053-0846  & 0.10$^{+0.26}_{-0.06}$ & 0.20$^{+0.13}_{-0.07}$ & $<$2.8 & 2.60$^{+0.36}_{-0.34}$ &
67.4$^{+46.6}_{-39.3}$ & =$\Gamma_1$ &  40.3(38)  \\

1123+4703  & 0.68$^{+0.70}_{-0.50}$ & $<$0.12 & $<$0.16 & 0.21$^{+0.39}_{-0.39}$ & 
& &  84.4 (89) \\

1346+6423$^1$  & 0.42$^{+0.22}_{-0.40}$ & $<$0.13 & 1 & 1.8 & & & 70.5 (43) \\

\hline
\hline
\multicolumn{8}{l}{*Best fit absorption is same as Galactic absorption.}\\
\multicolumn{8}{l}{``-'' denotes unconstrained parameter.}\\
\multicolumn{8}{l}{$^1\Gamma$ frozen to 1.8 to constrain spectral fit.}\\
 \end{tabular}
\end{table}

\begin{table}[h]
\small
\caption{\label{ftest_apec} F-test on \textit{APEC} fits}
\begin{tabular}{lcc}
\hline
\hline\noalign{\smallskip}
Target &  F$_{calc}$ & Probability\\
\hline\noalign{\smallskip}

0325-0608 & 8.83 & $>$0.9973 \\

1018+3722 & 8.53 & $>$0.9973 \\

1123+4703 & 7.59 & $>$0.9973 \\

1147+5226 & 8.63 & $>$0.9973 \\

\hline
\hline
 \end{tabular}
\end{table}

\begin{table}[h]
\caption{\label{Fe}Fe K$\alpha$}
\begin{tabular}{lccccc}
\hline
\hline\noalign{\smallskip}
Target &  Energy$^1$ & $\sigma$ & EW  & Flux & log L$_{Fe K\alpha}$\\
& keV & keV & keV &  10$^{-14}$ ergs cm$^{-2}$ s$^{-1}$ & ergs s$^{-1}$\\
\hline\noalign{\smallskip}

0053-0846$^3$  & 6.28 & 0.01 & $<$3.4 &  $<$8.5 & $<$40.83\\

0325-0608  & 6.19 & 0.01 & 0.22$^{+0.15}_{-0.15}$   & 2.12$^{+1.48}_{-1.47}$ & 40.73$^{+0.23}_{-0.54}$ \\

0800+2636$^2$ & 6.26$^{+0.03}_{-0.03}$ & $<$0.09 & 0.19$^{+0.26}_{-0.05}$  & 
8.04$^{+2.16}_{-2.03}$ & 41.11$^{+0.11}_{-0.13}$\\

0804+2345$^3$  & 6.22 & 0.01 & $<$1.5 &  $<$1.0               & $<$40.27\\

0824+2959$^2$ & 6.26$^{+0.24}_{-0.07}$ & 0.27$^{+0.33}_{-0.10}$ & 0.77$^{+0.74}_{-0.24}$ & 
11.1$^{+10.7}_{-3.45}$ & 41.19$^{+0.29}_{-0.16}$ \\

0959+1259$^2$ & 6.19$^{+0.25}_{-0.09}$ & - & 0.18$^{+0.09}_{-0.09}$   & 
1.40$^{+1.71}_{-0.86}$  & 40.56$^{+0.35}_{-0.41}$\\

1018+3722$^3$  & 6.10 & 0.01 & -  &  $<$1.51      & $<$40.92 \\

1111+0228$^3$  & 6.18 & 0.01 & $<$0.89 &  $<$2.1  & $<$40.76 \\

1123+4703$^3$  & 6.24 & 0.01 & $<$1.8  & $<$2.2   & $<$40.48\\

1136+5657$^3$  & 6.09 & 0.01 & $<$1.0  & $<$1.2   & $<$40.86\\

1147+5226  & 6.15 & 0.01 & 1.03$^{+0.44}_{-0.44}$   & 2.67$^{+1.13}_{-1.13}$   & 41.17$^{+0.15}_{-0.23}$\\

1157+5249  & 6.18 & 0.01 & 2.18$^{+0.66}_{-0.71}$   & 2.41$^{+0.72}_{-0.78}$   & 40.85$^{+0.11}_{-0.15}$\\

1218+4706  & 5.85 & 0.01 & 1.11$^{+0.66}_{-0.67}$   & 0.14$^{+0.08}_{-0.09}$ & 40.48$^{+0.20}_{-0.45}$ \\

1238+0927$^2$ & 5.90$^{+0.03}_{-0.03}$ & $<$0.09 & 0.17$^{+0.07}_{-0.06}$   
& 4.34$^{+1.78}_{-1.56}$ & 41.85$^{+0.15}_{-0.20}$ \\

1323+4318$^2$ & 6.18$^{+0.30}_{-0.12}$ & 0.23$^{+0.28}_{-0.15}$ & 0.76$^{+0.89}_{-0.43}$    
& 12.0$^{+14.0}_{-6.80}$  & 41.29$^{+0.33}_{-0.37}$ \\
           
1346+6423$^3$  & 6.25 & 0.01 & -  &  $<$7.1        & $<$40.96 \\

1437+3634$^3$  & 6.31 & 0.01 & -  & $<$5.4         & $<$40.36 \\

\hline
\hline
\multicolumn{6}{l}{$^1$Energy in observed frame.}\\
\multicolumn{6}{l}{$^2$Targets with the highest signal-to-noise for the Fe K$\alpha$ line detection.}\\
\multicolumn{6}{l}{$^3$Compton-thick candidates with coadded MOS spectra used in Figure \ref{ews}.}\\
\multicolumn{6}{l}{See Section 4.2 for details.}\\
\multicolumn{6}{l}{Upper limits on EW, flux and luminosity are 3$\sigma$ upper limits;}\\
\multicolumn{6}{l}{upper limits on $\sigma$ are 90\% upper limits.}\\
\multicolumn{6}{l}{``-'' denotes unconstrained parameter.}\\
\end{tabular}
\end{table}

\begin{table}[h]
\small
\caption{\label{disk}Double Powerlaw + Diskline Model Parameters}
\begin{tabular}{lcccccccr}
\hline
\hline\noalign{\smallskip}
Target & N$_H$ & $\Gamma_1$ & N$_H$ & $\Gamma_2$ & E & Inclination & EW & $\chi^2$ \\
&        10$^{22}$ cm$^{-2}$ &    &  10$^{22}$  cm$^{-2}$ & & & & keV & (DOF) \\
\hline\noalign{\smallskip}

0800+2636* & 0.03 & 1.29$^{+0.06}_{-0.07}$ & 1.11$^{+0.07}_{-0.08}$ & =$\Gamma_1$ &
6.33$^{+0.03}_{-0.03}$ & $<$15 & 0.25$^{+0.05}_{-0.08}$ & 501.4 (448)\\

0824+2959* & 0.03 & 2.74$^{+0.07}_{-0.07}$ & 22.0$^{+1.6}_{-1.6}$ & =$\Gamma_1$ & 6.23$^{+0.08}_{-0.07}$ &
36.2$^{+23.4}_{-9.1}$ & 0.97$^{+0.37}_{-0.21}$ & 224.3 (165) \\

0959+1259 & 0.77$^{+0.06}_{-0.06}$ & 1.88$^{+0.10}_{-0.08}$ & & & 
6.27$^{+0.03}_{-0.19}$ & - & 0.20$^{+0.09}_{-0.09}$ & 206.5 (267)\\

1238+0927 & 0.07$^{+0.06}_{-0.05}$ & 2.97$^{+0.47}_{-0.39}$ & 29.1$^{+4.0}_{-3.2}$ & =$\Gamma_1$ &
6.13$^{+0.07}_{-0.07}$ & $>62.1$ & 1.22$^{+0.35}_{-0.38}$ & 177.7 (124) \\

1323+4318 &  0.08$^{+0.34}_{-0.06}$ & 2.89$^{+2.42}_{-0.83}$ & 42.8$^{+14.2}_{-11.5}$ & =$\Gamma_1$ &
 6.17$^{+0.15}_{-0.13}$ & - & 1.08$^{+0.86}_{-0.62}$ & 58.1 (64) \\

\hline
\hline
\multicolumn{9}{l}{* Best-fit absorption same as Galactic absorption.}\\
\multicolumn{9}{l}{``-'' denotes unconstrained parameter.}\\
\multicolumn{9}{l}{The \textit{diskline} model does not provide a statistically significant improvement over the use of}\\
\multicolumn{9}{l}{a gaussian component.}
\end{tabular}
\end{table}

\clearpage

\begin{table}[h]
\small
\caption{\label{lumin}X-ray Luminosity values}
\begin{tabular}{lccc}
\hline
\hline\noalign{\smallskip}
Target   & log L$_{APEC}$ & log L$_{X}$ & log L$_X$/L$_{[OIII],corr}$\\
         & erg s$^{-1}$   & erg s$^{-1}$ \\
         & 0.5-2keV       & 2-10keV      \\
\hline\noalign{\smallskip}

0053-0846 &  39.9 & 41.0$^{+1.0}_{-0.8}$ &  -0.49 \\

0325-0608 &  41.1 & 42.6$^{+0.3}_{-0.3}$ &   0.57 \\

0800+2636 &  39.9 & 42.9$^{+0.3}_{-0.3}$ &   1.34  \\

0804+2345 &  40.0 & 40.7$^{+0.5}_{-0.7}$ &  -0.50 \\

0824+2959 &  40.0 & 42.1$^{+0.3}_{-0.3}$ &   0.56 \\

0959+1259 &  39.8 & 42.5$^{+0.3}_{-0.3}$ &   0.83  \\

1018+3722 &  40.8 & 40.2$^{+0.3}_{-0.3}$ &  -1.40 \\

1111+0228 &  39.4 & 41.5$^{+0.4}_{-0.3}$ &   0.21 \\

1123+4703 &  40.1 & 41.3$^{+0.3}_{-0.3}$ &  0.11 \\

1136+5657 &  40.5 & 41.6$^{+0.6}_{-0.5}$ &  -0.72  \\

1147+5226 &  41.1 & 42.2$^{+0.5}_{-0.4}$ &  0.14  \\

1157+5249 &  40.5 & 41.5$^{+0.4}_{-0.5}$ &   0.07  \\

1218+4706 &  41.1 & 42.5$^{+0.4}_{-0.6}$ &   0.03 \\

1238+0927 &  41.0 & 43.2$^{+0.3}_{-0.3}$ &   0.87 \\

1323+4318 &  39.9 & 42.1$^{+0.3}_{-0.5}$ &   0.47 \\

1346+6423 &  40.2 & 40.3$^{+0.3}_{-0.3}$ &  -0.90 \\

1437+3634 &  39.5 & 39.6$^{+0.4}_{-0.3}$ &  -1.15 \\

\hline
\hline
\end{tabular}
\end{table}

\begin{table}[h]
\small
\caption{\label{spitzer}SDSS and Spitzer Data}
\begin{tabular}{lcccccccc}
\hline
\hline\noalign{\smallskip}
Target & SFR & L$_{[OIII],obs}$ &  L$_{[OIII],corr}$ &  L$_{[OIV]}$ & L$_{MIR}$ &  M$_{BH}$ & SFR/M* &  L$_{[OIII]}$/M$_{BH}$ \\
       & M$_{\sun}$/yr  & log L$_{\sun}$ & log L$_{\sun}$ & log  L$_{\sun}$ & log  L$_{\sun}$ & log M$_{\sun}$ 
& log yr$^{-1}$ & log L$_{\sun}$/M$_{\sun}$  \\
\hline\noalign{\smallskip}

0053-0846 & 1.04 & 7.14 & 7.86 & 7.71 & 10.26 & 7.11 & -10.65  & 0.76 \\

0325-0608 & 12.0 & 7.74 & 8.40 & 7.69 & 10.67 & 7.80 & -10.06  & 0.60 \\

0800+2636 & 0.74 & 7.29 & 7.92 & 7.64 & 10.31 & 7.74 & -10.98  & 0.18 \\

0804+2345 & 0.52 & 7.43 & 7.62 & 7.65 &  9.29 & 7.25 & -10.83  & 0.37 \\

0824+2959 & 0.64 & 7.53 & 7.95 & 7.03 & 10.56 & 6.99 & -10.06  & 0.96 \\

0959+1259 & 0.85 & 7.78 & 8.04 & 7.44 &  9.78 & 6.78 &  -9.86  & 1.26 \\

1018+3722 & 1.33 & 7.79 & 8.05 & 7.56 & 10.26 & 7.14 &  -9.88  & 0.91 \\

1111+0228 & 0.47 & 7.49 & 7.71 & 7.27 &  9.92 & 7.33 & -10.76  & 0.38 \\

1123+4703 & 0.57 & 7.24 & 7.57 & 6.86 &  9.13 & 7.29 & -10.96  & 0.28 \\

1136+5657 & 5.11 & 7.89 & 8.70 & 8.35 & 11.20 & 7.99 & -10.27  & 0.71 \\

1147+5226 & 2.88 & 7.80 & 8.43 & 7.60 & 10.69 & 7.85 & -10.12  & 0.57 \\

1157+5249 & 0.60 & 7.58 & 7.83 & 7.57 & 10.26 & 7.17 & -10.76  & 0.67 \\

1218+4706 & ... & 8.46 & 8.86 & 8.30 & 10.77 & 8.17 &  ...    & 0.69 \\

1238+0927 & 6.30 & 8.39 & 8.74 & 8.37 & 11.01 & 8.40 & -10.35  & 0.34 \\

1323+4318 & 2.17 & 7.27 & 8.03 & 7.82 & 10.28 & 7.65 & -10.54  & 0.38 \\

1346+6423 & 0.36 & 7.16 & 7.60 & 6.80 &  9.55 & 6.81 & -10.27  & 0.79 \\

1437+3634 & 0.05 & 6.85 & 7.13 & 6.90 &  9.05 & 7.44 & -12.06  & -0.31 \\

\hline
\hline
\end{tabular}
\end{table}

\begin{table}[h]
\small
\caption{\label{gamma}Simulated N$_H$ (10$^{22}$ cm$^{-2}$) for $\Gamma$=1.8}
\begin{tabular}{lrrrr}
\hline
\hline\noalign{\smallskip}
Target & CF=0.99 & CF=0.98 & CF=0.97 & CF=0.96\\
\hline\noalign{\smallskip}

0053-0846 & 64  & 104 & 150 & 201 \\

0325-0608 & 2.5 & 2.6 & 2.8 & 2.9 \\ 

0800+2636 & 1.2 & 1.2 & 1.2 & 1.3 \\

0804+2345 & 205 & 356 & 461 & 566 \\

0824+2959 & 30  & 39  & 51  & 80  \\

0959+1259 & 11  & 13  & 16  & 20  \\

1018+3722 & 67  & 128 & 192 & 252 \\

1111+0228 & 84  & 157 & 225 & 300 \\

1123+4703 & 133 & 228 & 325 & 398 \\

1136+5657 & 23  & 23  & 24  & 258 \\

1147+5226 & 39  & 48  & 59  & 86  \\

1157+5249 & 76  & 150 & 219 & 294 \\

1218+4706 & 111 & 202 & 297 & 364 \\ 

1238+0927 & 19  & 21  & 24  & 27  \\

1323+4318 & 26  & 33  & 46  & 69  \\

1346+6423 & 228 & 385 & 498 & 607 \\

1437+3634 & 108 & 199 & 295 & 362 \\

\hline
\hline
 \end{tabular}
\end{table}

\begin{table}[h]
\small
\caption{\label{cf}Simulated N$_H$ (10$^{22}$ cm$^{-2}$) for CF=0.99}
\begin{tabular}{lrrrrr}
\hline
\hline\noalign{\smallskip}
Target & $\Gamma$=1.6 & $\Gamma$=1.7 & $\Gamma$=1.8 & $\Gamma$=1.9 & $\Gamma$=2.0\\
\hline\noalign{\smallskip}

0053-0846 & 65  & 65  & 64  & 64  & 63  \\

0325-0608 & 2.5 & 2.5 & 2.5 & 2.6 & 2.6 \\

0800+2636 & 1.1 & 1.1 & 1.2 & 1.2 & 1.2 \\

0804+2345 & 202 & 204 & 205 & 207 & 208 \\

0824+2959 & 31  & 30  & 30  & 30  & 29  \\

0959+1259 & 11  & 11  & 11  & 11  & 11  \\

1018+3722 & 67  & 67  & 67  & 67  & 66  \\

1111+0228 & 83  & 83  & 84  & 84  & 84  \\

1123+4703 & 132 & 133 & 133 & 134 & 134 \\

1136+5657 & 25  & 24  & 23  & 22  & 20  \\

1147+5226 & 40  & 40  & 39  & 38  & 37  \\

1157+5249 & 76  & 76  & 76  & 78  & 78  \\

1218+4706 & 111 & 111 & 111 & 112 & 111 \\

1238+0927 & 20  & 19  & 19  & 18  & 18  \\

1323+4318 & 27  & 26  & 26  & 26  & 26  \\

1346+6423 & 224 & 227 & 228 & 230 & 233 \\

1437+3634 & 106 & 108 & 108 & 111 & 112 \\

\hline
\hline
 \end{tabular}
\end{table}

\begin{table}[h]
\small
\caption{\label{plcabs}N$_H$ (10$^{22}$ cm$^{-2}$) values using \textit{plcabs}}
\begin{tabular}{lrr}
\hline
\hline\noalign{\smallskip}
Target & N$_H$ & N$_H$ \\
       & fitted & simulated \\ 
\hline\noalign{\smallskip}

0053-0846 & 0.05 & 49 \\

0325-0608 & 0.04 & 2.7 \\

0800+2636 & 1.1  & 1.2 \\

0804+2345 & 68  & 87 \\

0824+2959 & 25  & 27 \\

0959+1259 & 0.89 & 11 \\

1018+3722 & 0.23 & 46 \\

1111+0228 & 5.7  & 57 \\

1123+4703 & 2.2  & 70 \\

1136+5657 & 63   & 22 \\

1147+5226 & 32   & 37 \\

1157+5249 & 138    & 51 \\

1218+4706 & 120 & 72 \\

1238+0927 & 41    & 20 \\

1323+4318 & 50  & 23 \\

1346+6423* & ... & 93 \\

1437+3634 & 14  & 60 \\

\hline
\hline
\multicolumn{3}{l}{*Due to low S/N, \textit{plcabs} could not}\\
\multicolumn{3}{l}{adequately fit the spectrum.}\\
 \end{tabular}
\end{table}

\end{document}